\documentclass[nofootinbib,floatfix,
superscriptaddress]{revtex4}        
\usepackage{graphicx}
\usepackage{epsfig}
\usepackage{bm}
\usepackage[T1]{fontenc}
\usepackage[latin9]{inputenc}
\usepackage{amssymb}
\usepackage{float}
\usepackage{amsmath}
\usepackage{dcolumn}
\usepackage{cancel}
\usepackage[colorlinks]{hyperref}
\usepackage[usenames,dvipsnames]{color}
\hypersetup{
     breaklinks=true,
    pdfstartview={FitH},    
    colorlinks=true,       
    linkcolor=blue,          
    citecolor=red,        
    filecolor=magenta,      
    urlcolor=blue,           
    anchorcolor=green,      
    linktocpage=true
}

\newcommand{\be}{\begin{equation}}
\newcommand{\ee}{\end{equation}}
\newcommand{\De}{\Delta}

\begin{document}

\title{Generalized interacting Barrow Holographic Dark Energy: cosmological predictions and thermodynamic considerations}

\author{G.~G.~Luciano}
\email{giuseppegaetano.luciano@udl.cat (corresponding author)}
\affiliation{Applied Physics Section of Environmental Science Department, Universitat de Lleida, Av. Jaume
II, 69, 25001 Lleida, Catalonia, Spain}

\author{J.~Gin\' e}
\email{gine@matematica.udl.cat}
\affiliation{Departament de Matemàtica, Universitat de Lleida, Av. Jaume II, 69, 25001 Lleida, Spain}

\date{\today}

\begin{abstract}
We construct a generalized interacting model of Barrow Holographic Dark Energy (BHDE) with infrared cutoff being given by 
the Hubble horizon. We analyze the cosmological evolution of a 
flat Friedmann-Lema\^itre-Robertson-Walker Universe filled by pressureless dark matter, BHDE and radiation fluid. The interaction  
between the dark sectors of the cosmos is assumed 
of non-gravitational origin and satisfying the second law of thermodynamics and Le Chatelier-Braun principle.  
We study the behavior of various model parameters, such
as the BHDE density parameter, the equation of state
parameter, the deceleration parameter, the jerk parameter and
the square of sound speed. 
We show that our model satisfactorily retraces the thermal history of the Universe and is consistent with current observations for certain values of parameters, providing an eligible candidate to describe dark energy. We
finally explore the thermodynamics of 
our framework, with special focus on the validity of the generalized second law.
\end{abstract}
 \maketitle

\section{Introduction}
\label{Intro}
It is nowadays established experimentally that our Universe
is experiencing an accelerated expansion~\cite{Supern,Supernbis,Supernter,Supernquar,Supernquin}. Despite
the intensive theoretical and observational studies, the origin of this phenomenon is yet to be understood, leaving room for disparate explanations. Among the various scenarios, 
a plausible mechanism is the existence 
of an exotic form of energy with large negative pressure - the Dark Energy (DE) - which should affect the Universe on large scales. 
However, the true nature of this entity remains unknown,
and a vast number of theoretical models have been proposed to
shed light on its properties~\cite{Peebles:1987ek,Ratra:1987rm,Wetterich:1987fm,Frieman:1995pm,Turner:1997npq,Caldwell:1997ii,Armendariz-Picon:2000nqq,Amendola:1999er,Armendariz-Picon:2000ulo,Kamenshchik:2001cp,Caldwell:1999ew,Caldwell:2003vq,Nojiri:2003vn,Feng:2004ad,Guo:2004fq,Elizalde:2004mq,Nojiri:2005sx,Deffayet:2001pu,Sen:2002in,Zhangov}.

In this context, a valuable approach is the so called Holographic Dark Energy (HDE) model~\cite{Cohen:1998zx,Horava:2000tb,Thomas:2002pq,Li:2004rb,Hsu:2004ri,Huang:2004ai,Wang:2005ph,Setare:2006sv,Granda,Sheykhi:2011cn,Bamba:2012cp,Ghaffari:2014pxa,Wang:2016och,Moradpour:2020dfm}, which is based on the holographic principle and the use of Bekenstein-Hawking (BH) area law~\cite{Beke} for the entropy on the Universe horizon. 
Cosmological applications of HDE have been extensively considered
in literature~\cite{Enqvist:2004ny,Setare:2007hq,Sriva}. 
Nevertheless, the shortcomings of this model in describing the
history of the Universe have motivated tentative changes over the years.
For instance, several extensions have been developed to accommodate non-additive~\cite{Tavayef:2018xwx,Saridakis:2018unr,Nojiri:2019skr}, relativistic~\cite{Drepanou:2021jiv,Hernand} and quantum-like~\cite{BarSar,Dabrowski:2020atl,Sheykhi:2021fwh,Adhikary:2021xym,Nojiri:2021jxf} corrections to BH entropy. The ensuing frameworks are referred to as Tsallis~\cite{Tavayef:2018xwx,Saridakis:2018unr,Nojiri:2019skr,Luciano:2022ely}, Kaniadakis~\cite{Drepanou:2021jiv,Hernand,ModKa} and Barrow~\cite{BarSar,Dabrowski:2020atl,Sheykhi:2021fwh,Adhikary:2021xym,Nojiri:2021jxf,LucianoBarrow,Anagnostopoulos:2020ctz,GhaffariLuc,LucTherm} HDE, respectively, 
and have been shown to well fit experimental data
for certain values of the model parameters.

In various cosmological scenarios, it is typically assumed
that dark energy interacts with dark matter (DM)~\cite{ReviewDM,Addazi}, the existence
of which is invoked to explain the observed flatness of the rotation curves of spiral galaxies. Besides, such an interaction might 
play a non-trivial r\^ole in solving some open problems
in modern Cosmology, such as the coincidence problem~\cite{Amendola,Bolotin:2013jpa} 
and the current tension on the local value of the Hubble constant~\cite{DiVale,Kumar:2017dnp}. Therefore, cosmological models
involving interactions between the dark sectors of the cosmos
turn out to be particularly compelling and
deserve special attention.

Along this line, interacting models of HDE based on non-extensive Tsallis entropy have been investigated in a variety of contexts, ranging from higher dimensional Cosmology~\cite{Saha}, to the Brans-Dicke Cosmology~\cite{BDCosmo} and DGP brane-world~\cite{DGP}. Recently, a theoretical model of generalized interacting Tsallis HDE has been studied in~\cite{Mamon:2020wnh} along with its thermodynamic implications. Inspired by this analysis, the dynamics of the interacting
HDE based on Barrow entropy (Barrow Holographic Dark Energy, BHDE) has been proposed in~\cite{Mamon:2020spa} for a spatially flat Friedmann-Lema\^itre-Robertson-Walker (FLRW) Universe composed of 
pressureless DM and BHDE. 

Starting from the above premises, in this manuscript we 
explore the consequences of the interacting BHDE model 
in a more general scenario. In a recent work by one of the authors~\cite{LucTherm},  the cosmic evolution and thermal stability of BHDE have been studied in nonflat Friedmann-Robertson-Walker geometry by considering a particular
form for the interaction and neglecting the radiation content of the Universe. Here we extend the above analysis to the case of the most
general interaction satisfying Le Chatelier-Braun principle
and also take account of radiation fluid effects, which 
are particularly relevant in the early stages of Universe existence.
In this sense, our work also extends the results of~\cite{Mamon:2020spa}, where a Universe filled up only with dark matter and energy
has been considered. In this background, 
we analyze the behavior of the BHDE density parameter, the deceleration parameter, the jerk parameter and the
BHDE equation of state parameter, as well as the
classical stability and thermodynamic nature of our model, 
showing that our generalized scenario 
has a phenomenology richer than the one of~\cite{Mamon:2020spa,LucTherm}. This clearly offer fresh insights toward a more comprehensive understanding
of the intimate nature of dark energy. 
We also prove that the present framework is consistent with recent observations,  
thus providing a good candidate for the description of DE.

The remainder of the work is organized as follows: in the next Section 
we describe the interacting BHDE model. In particular, we study the cosmological evolution of some characteristic parameters of the model for a Universe filled by pressureless DM and BHDE. In Sec.~\ref{Rad} we 
extend this analysis to the case where the
radiation fluid is also present.  
Sec.~\ref{Thermo} is devoted to investigate the thermodynamic implications of this generalized BHDE model. Conclusions and outlook are finally summarized in Sec.~\ref{Concl}. Throughout the entire manuscript, we work in units where $\hbar=k_B=c=1$, while we keep the gravitational constant $G$ explicit. Furthermore, we use 
the mostly negative signature for the metric. 

\section{Interacting BHDE model with Hubble scale as IR cutoff}
\label{InterBHDE}
Let us start by describing the framework of the interacting BHDE model and its cosmological implications. 
BHDE is based upon the holographic principle with the
following modified entropy-area law for the Universe horizon~\cite{BarSar}
\be
\label{BE}
S_\De\,=\,\left(\frac{A}{A_0}\right)^{1+\Delta/2}\,, 
\ee
where $A$ is the area enclosed by the horizon surface
and $A_0=4G$ the Planck area. Although this relation 
was originally proposed by Barrow to accommodate
quantum-gravitational effects on the horizon of black holes~\cite{BarrowBH}, 
it can also be applied in the cosmological framework
based on the gravity-thermodynamic conjecture~\cite{BarSar,Dabrowski:2020atl,Sheykhi:2021fwh,Adhikary:2021xym,Nojiri:2021jxf}. 
In this way, extra corrections appear in the
Standard Model of Cosmology, namely on the Friedmann 
equations. Such terms are
parameterized by the exponent $0\le\Delta\le1$, 
where $\Delta=0$ gives the standard Bekenstein-Hawking
limit, while $\Delta=1$ corresponds to the maximal horizon
deformation. In passing, we mention that constraints on $\Delta$ 
have been derived in~\cite{Anagnostopoulos:2020ctz,Leon:2021wyx,Barrow:2020kug,Jusufi:2021fek,Dabrowski:2020atl,Saridakis:2020cqq,Luciano:2022pzg,Vagnozzi:2022moj}. Furthermore, the possibility of a running $\Delta$ has been discussed in~\cite{DiGennaro:2022ykp}.

We now observe that the usage of Barrow entropy~\eqref{BE} leads
to the following holographic dark energy density~\cite{BarSar}
\be
\label{rhoD}
\rho_D=CL^{\Delta-2}\,,
\ee
where $C$ is an unknown parameter with dimensions $[L]^{-2-\Delta}$ 
and $L$ is the characteristic size of the system. 
It is worth noticing that, for $\Delta=0$, the above relation
reduces to the standard HDE~\cite{Cohen:1998zx,Horava:2000tb,Thomas:2002pq,Li:2004rb,Hsu:2004ri,Huang:2004ai,Wang:2005ph,Setare:2006sv,Granda,Sheykhi:2011cn,Bamba:2012cp,Ghaffari:2014pxa,Wang:2016och,Moradpour:2020dfm}, provided that $C=3c^2m_p^2$, where $m_p$ is the reduced Planck mass and $c$
a dimensionless constant (not to be confused with the speed of light). 
Conversely, in the case where
deformations effects are taken into account (i.e. $\Delta\neq0$), BHDE
gives rise to a more general framework than HDE~\cite{BarSar,Dabrowski:2020atl,Sheykhi:2021fwh,Adhikary:2021xym,Nojiri:2021jxf}. 

By considering the Hubble horizon $L=H^{-1}$
as the IR cutoff, the energy density of BHDE can be rewritten as\footnote{For some arguments on the use of Hubble horizon as the IR cutoff, see~\cite{Mamon:2020spa}. For the sake of completeness, we mention that other possible choices are the particle horizon, the future event horizon, the Granda-Oliveros cutoff or combination thereof.} 
\be
\label{IRcut}
\rho_D=CH^{2-\Delta}\,.
\ee
This relation provides
the crucial ingredient of our next analysis. 

Let us assume that the Universe is homogenous, isotropic and
spatially flat on large scales, so that it can be
described by the standard Friedmann-Lema\^itre-Robertson-Walker (FLRW) metric
\be
\label{FLRW}
ds^2=dt^2-a^2(t)\left(dr^2+r^2d\Omega^2\right),
\ee
where $a(t)$ is the scale factor depending on the cosmic time $t$
and $r,\Omega$ are the radial and angular coordinates, respectively. 
Under the further assumption that the Universe 
be filled by pressureless DM and BHDE
of densities $\rho_m$ and $\rho_D$, respectively, 
Friedmann equations take the standard form
\begin{eqnarray}
\label{F1}
H^2&=&\frac{1}{3m_p^2}\left(\rho_m+\rho_D\right),\\[2mm]
\dot H&=&-\frac{1}{2m_p^2}\left(\rho_m+\rho_D+p_D\right),
\label{F2}
\end{eqnarray}
where $H=\dot a/a$ is the Hubble parameter and
the overdot denotes derivative respect to $t$. The 
pressure of BHDE has been indicated by $p_D$ and is related
to the energy density $\rho_D$ via the equation of state (EoS) 
\be
\label{eos}
\omega_D\equiv\frac{p_D}{\rho_D}\,,
\ee 
where $\omega_D$ is the so called
EoS parameter.

We can also introduce the fractional energy density parameters
of DM ($\Omega_m$) and BHDE ($\Omega_D$) as
\begin{eqnarray}
\label{Omm}
\Omega_m&=&\frac{\rho_m}{\rho_c}\,,
\\[2mm]
\Omega_D&=&\frac{\rho_D}{\rho_c}=\frac{C}{3m_p^2}H^{-\Delta}\,,
\label{Omd}
\end{eqnarray}
where 
\be
\rho_c=3m_p^2H^2
\ee
is the critical energy density. 
In this way, Eq.~\eqref{F1} can be equivalently rewritten as $\Omega_m+\Omega_D=1$. 
Moreover, we define the ratio of the energy densities as
$r=\rho_m/\rho_D=\Omega_m/\Omega_D$. 

Let us now assume that DM and BHDE nontrivially interact with
each other. Denoting by $Q$ the rate of energy exchange, 
the conservation equation for the dark sectors of the Universe can be
written as follows
\begin{subequations}
\label{Q1}
\begin{equation}
\dot\rho_m+3H\rho_m=Q\,,
\ee
\be
\dot\rho_D+3H\left(1+\omega_D\right)\rho_D=-Q\,.
\ee
\end{subequations}
Some comments on the sign and functional form of the interaction
$Q$ are in order here. First, we notice that for $Q>0$, 
energy flows from BHDE to DM, while the opposite happens
for $Q<0$. Second, 
while many interactions have been proposed to describe
the dynamics of the Universe, the exact
form of $Q$ is still unknown (see~\cite{Q1,Q2,Q3,Q4} for more details).
Following~\cite{Mamon:2020spa,Mamon:2020wnh}, hereafter
we assume
\be
\label{Q}
Q=3H\left(b_1^2\rho_m+b_2^2\rho_D\right),
\ee
where the model parameters $b_1$ and $b_2$ are
dimensionless constants. This form of $Q$
is largely used in the literature for different choices
of $b_1$ and $b_2$. Since in our setting the coefficients 
$b_1^2$ and $b_2^2$  
are always positive, we have $Q>0$, consistently with 
the validity of the second law of thermodynamics
and Le Chatelier-Braun principle~\cite{Q2}. We also
require $b_1^2\neq b_2^2$ and $b_1^2,b_2^2\ll1$
for observational reasons~\cite{Chimento}. 

Now, by taking the time derivative of Eq.~\eqref{F1}
combined with Eqs.~\eqref{Q1}, we obtain 
after some algebra
\be
\label{dotH}
\frac{\dot H}{H^2}=-\frac{3}{2}\Omega_D\left(1+\omega_D+r\right). 
\ee
In a similar fashion, the derivative of Eq.~\eqref{IRcut}
leads to 
\be
\label{omegaD}
\omega_D=\frac{-\frac{\Delta}{2}-\frac{b_1^2}{\Omega_D}+b_1^2-b_2^2}{1-\left(1-\frac{\Delta}{2}\right)\Omega_D}\,.
\ee
Notice that this function is divergent for
$\Omega_D={(1-\Delta/2)}^{-1}$. However, since by definition $\Omega_D<1$, such condition 
cannot be satisfied by any $0\le\Delta\le1$. 

The dynamical evolution of the BHDE density parameter
$\Omega_D$ can be obtained by deriving Eq.~\eqref{Omd}
respect to $t$ and then resorting to Eqs.~\eqref{dotH} and~\eqref{omegaD}. We get the following differential equation
\be
\label{Omp}
\Omega'_D=\frac{3\Delta\Omega_D}{2}\left[\frac{1-\Omega_D-b_1^2+\left(b_1^2-b_2^2\right)\Omega_D}{1-\left(1-\frac{\Delta}{2}\right)\Omega_D}\right],
\ee
where the prime denotes derivative respect to $\log a$, i.e.
$\Omega'_D=\frac{d\Omega_D}{d(\log a)}$. For our next purposes, 
we also remind that 
\be
\frac{1}{a}=\frac{1+z}{a_0}\,,
\ee
where $z$ denotes the redshift and $a_0$ is the present value
of the scale factor. Here, we take $a_0=1$, so that $a^{-1}=1+z$. 

In order to determine the behavior of BHDE density parameter, 
we solve Eq.~\eqref{Omp} by numerical integration\footnote{More details toward the analytic resolution of Eq.~\eqref{Omp} can be found
in the Appendix.}. 
Results are shown in Fig.~\ref{Fig1} and Fig.~\ref{Fig2}
for different values of the interaction terms $b_1^2, b_2^2$ and Barrow
parameter $\Delta$, respectively. 
From these plots, it is evident that $\Omega_D$ increases monotonically from zero at early times to unity in the far future (i.e. $z\rightarrow-1$), indicating an evolution toward a completely dark energy dominated epoch predicted by this model.

\begin{figure}
    \centering
    \begin{minipage}{0.45\textwidth}
        \centering
\includegraphics[width=8.5cm]{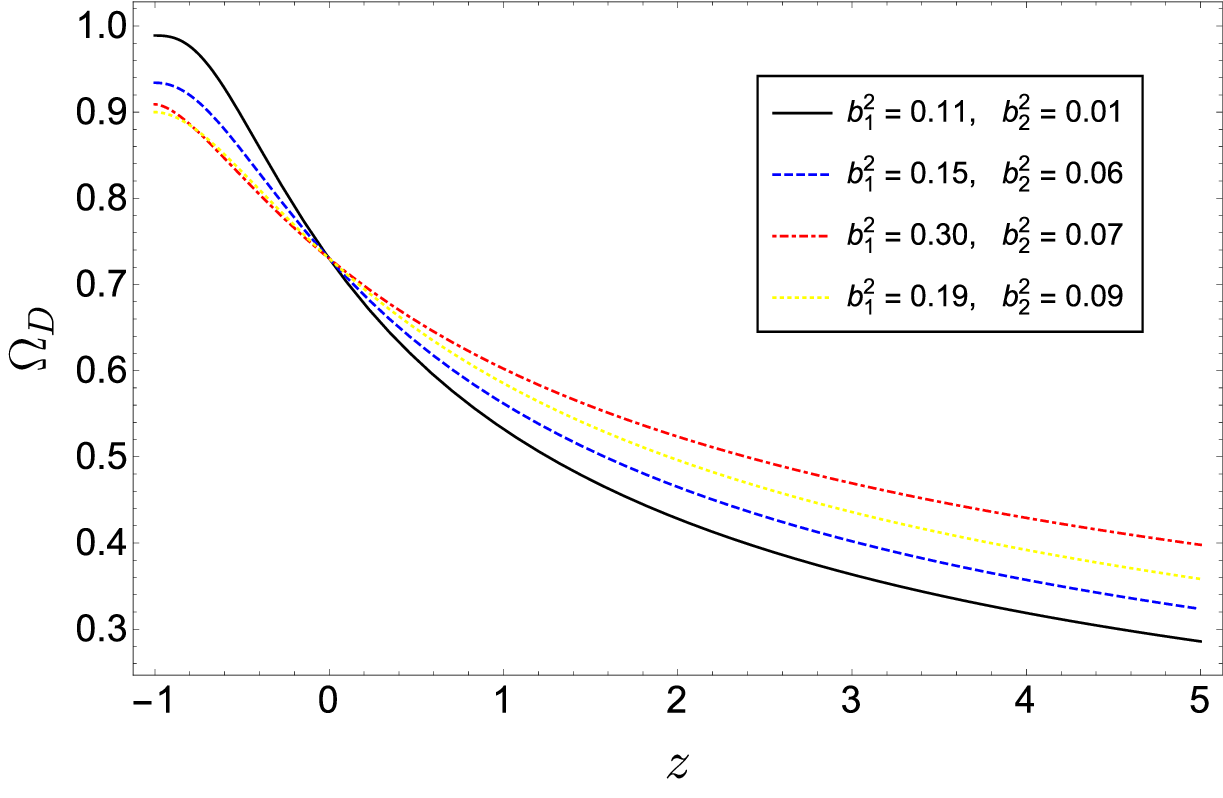}
 \caption{The evolution of $\Omega_D$ versus $z$ for different values
 of $b_1^2$ and $b_2^2$. We set $\Delta=0.5$ and $\Omega_D^0=0.73$ as initial condition.}
 \label{Fig1}
    \end{minipage}\hfill
  \begin{minipage}{0.45\textwidth}
        \centering
  \hspace{-0.5cm}\includegraphics[width=8.5cm]{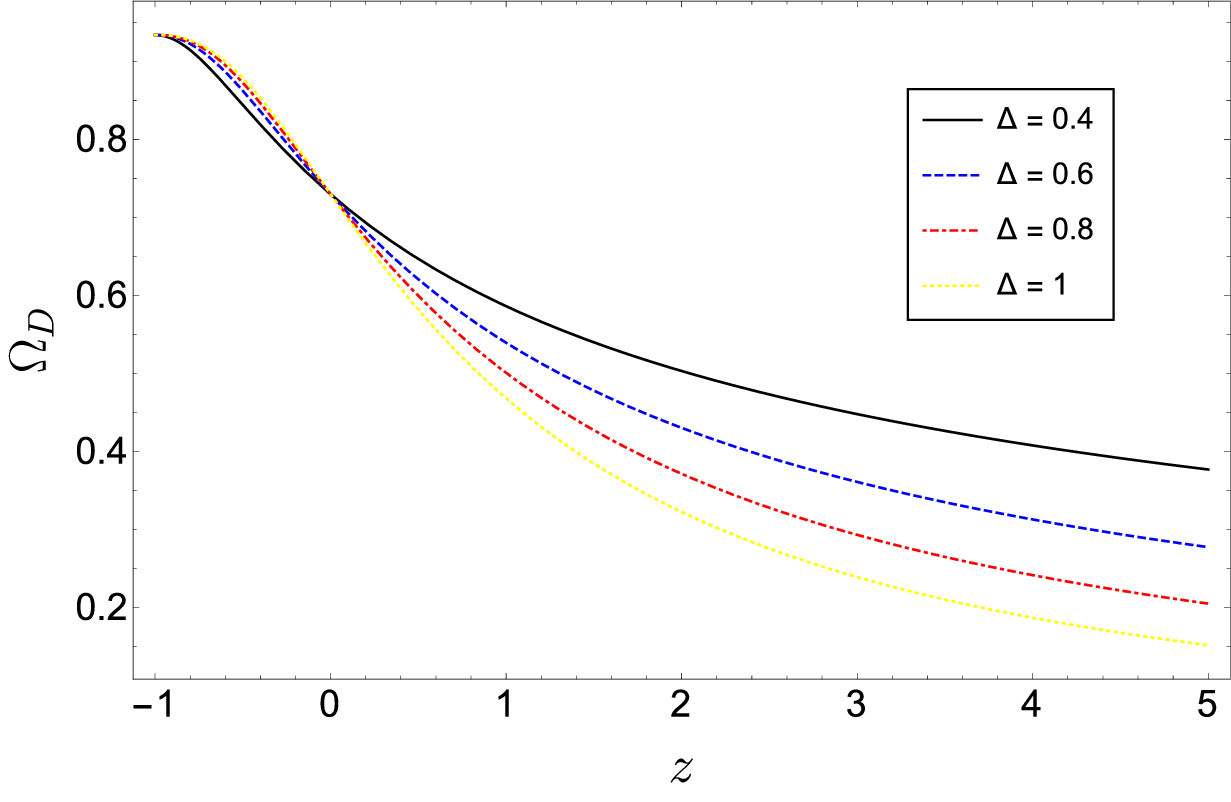}
 \caption{The evolution of $\Omega_D$ versus $z$ for different values
 of $\Delta$. We set $b_1^2=0.15$, $b_2^2=0.06$ and $\Omega_D^0=0.73$ as initial condition.}
 \label{Fig2}
    \end{minipage}
\end{figure}

%
%

\begin{figure}
    \centering
    \begin{minipage}{0.45\textwidth}
        \centering
\includegraphics[width=8.5cm]{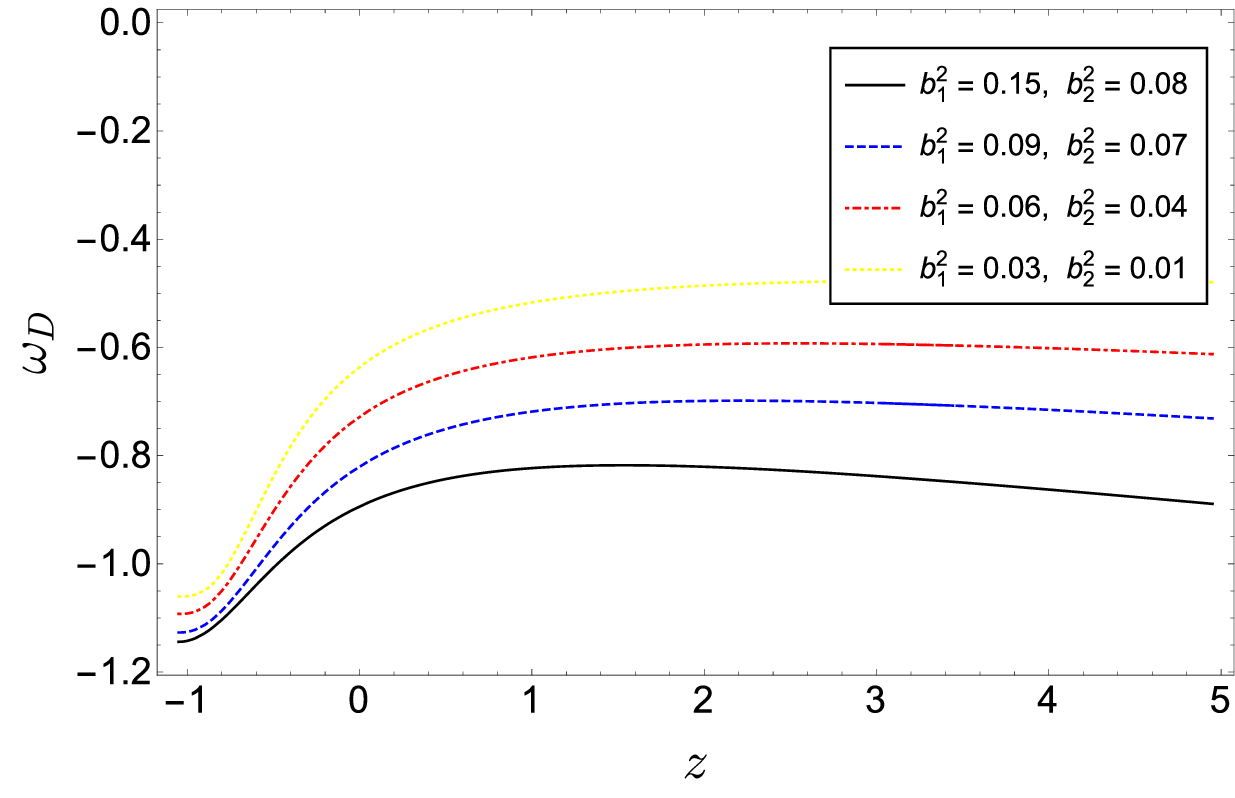}
 \caption{The evolution of $\omega_D$ versus $z$ for different values
 of $b_1^2$ and $b_2^2$. We set $\Delta=0.5$ and $\Omega_D^0=0.73$ as initial condition.}
 \label{Fig3}
     \end{minipage}\hfill
    \begin{minipage}{0.45\textwidth}
        \centering
\hspace{-0.5cm}\includegraphics[width=8.5cm]{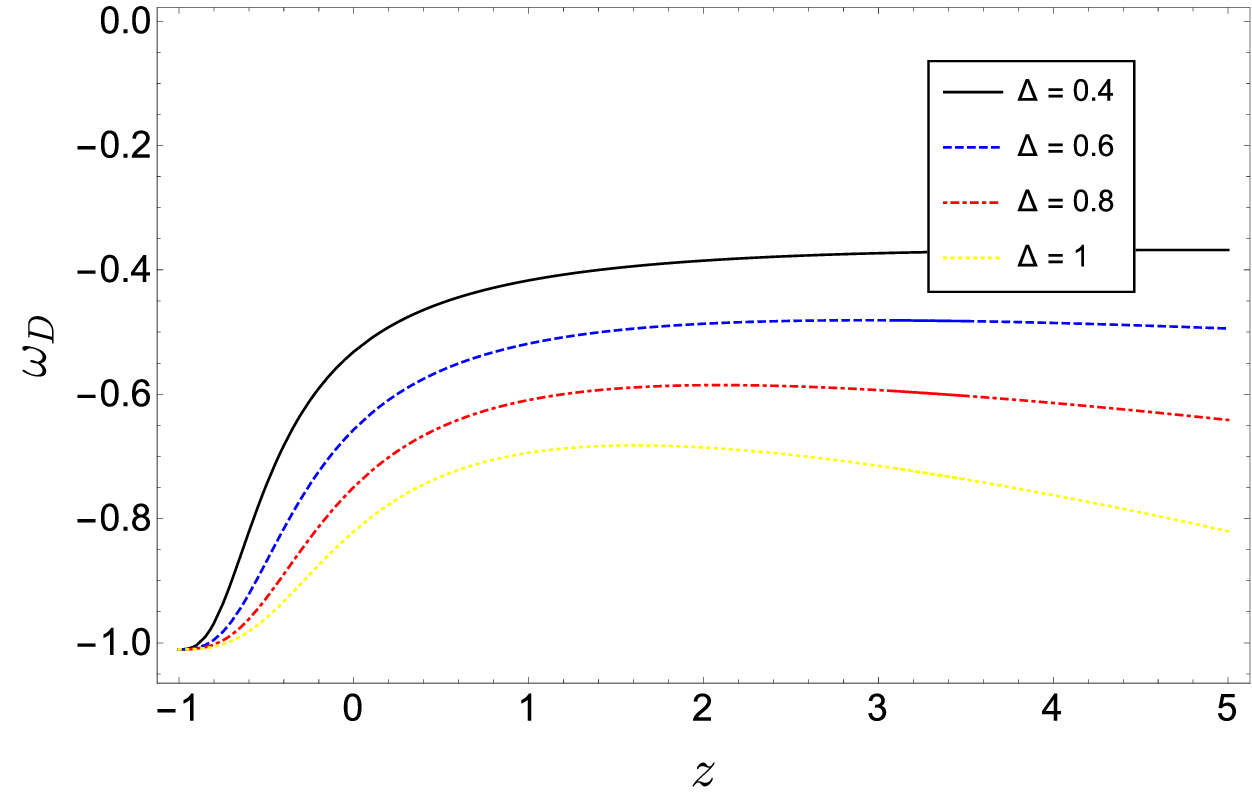}
 \caption{The evolution of $\omega_D$ versus $z$ for different values
 of $\Delta$. We set $b_1^2=0.03$, $b_2^2=0.01$ and $\Omega_D^0=0.73$ as initial condition.}
 \label{Fig4}
    \end{minipage}
\end{figure}

From Eq.~\eqref{omegaD}, we can infer the dynamics of the EoS
parameter $\omega_D$. Specifically, from Fig.~\ref{Fig3}
we see that $\omega_D$ always lies in the quintessence regime ($-1<\omega_D<-1/3$) at present (i.e. $z=0$) for the considered values of $b_1^2, b_2^2$. However, it first approaches the cosmological constant behavior ($\omega_D=-1$) and then
crosses the phantom line ($\omega_D<-1$) in the far future (i.e. $z\rightarrow-1$) for increasing $b_1^2, b_2^2$. The same profile
is exhibited by varying the Barrow parameter $\Delta$ and keeping $b_1^2,b_2^2$ fixed (see Fig.~\ref{Fig4}). To check the observational consistency of our model, we can also estimate
the value of the EoS parameter at present epoch. From Fig.~\ref{Fig3}, 
it is found to satisfy $-0.90\le\omega_{D_0}\le-0.64$, while Fig.~\ref{Fig4}
gives $-0.82\le\omega_{D_0}\le-0.53$. Notice that the first interval
overlaps  with the recent 
observational constraint $-1.38<\omega_0<-0.89$ obtained from Planck+WP+BAO measurements~\cite{Planck}.

%
%

\begin{figure}
    \centering
    \begin{minipage}{0.45\textwidth}
        \centering
\includegraphics[width=8.5cm]{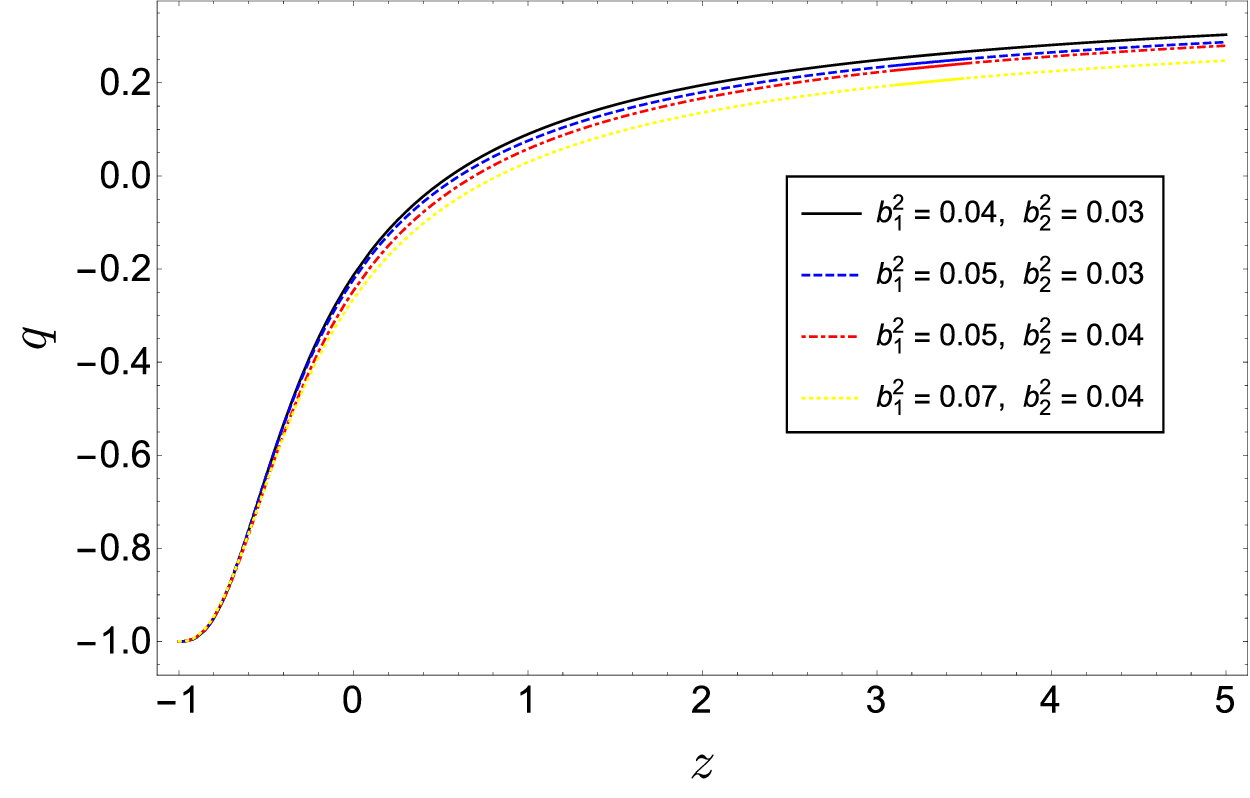}
 \caption{The evolution of $q$ versus $z$ for different values
of $b_1^2$ and $b_2^2$. We set $\Delta=0.5$ and $\Omega_D^0=0.73$ as initial condition.}
 \label{Fig5}
     \end{minipage}\hfill
    \begin{minipage}{0.45\textwidth}
        \centering
\hspace{-0.5cm}\includegraphics[width=8.5cm]{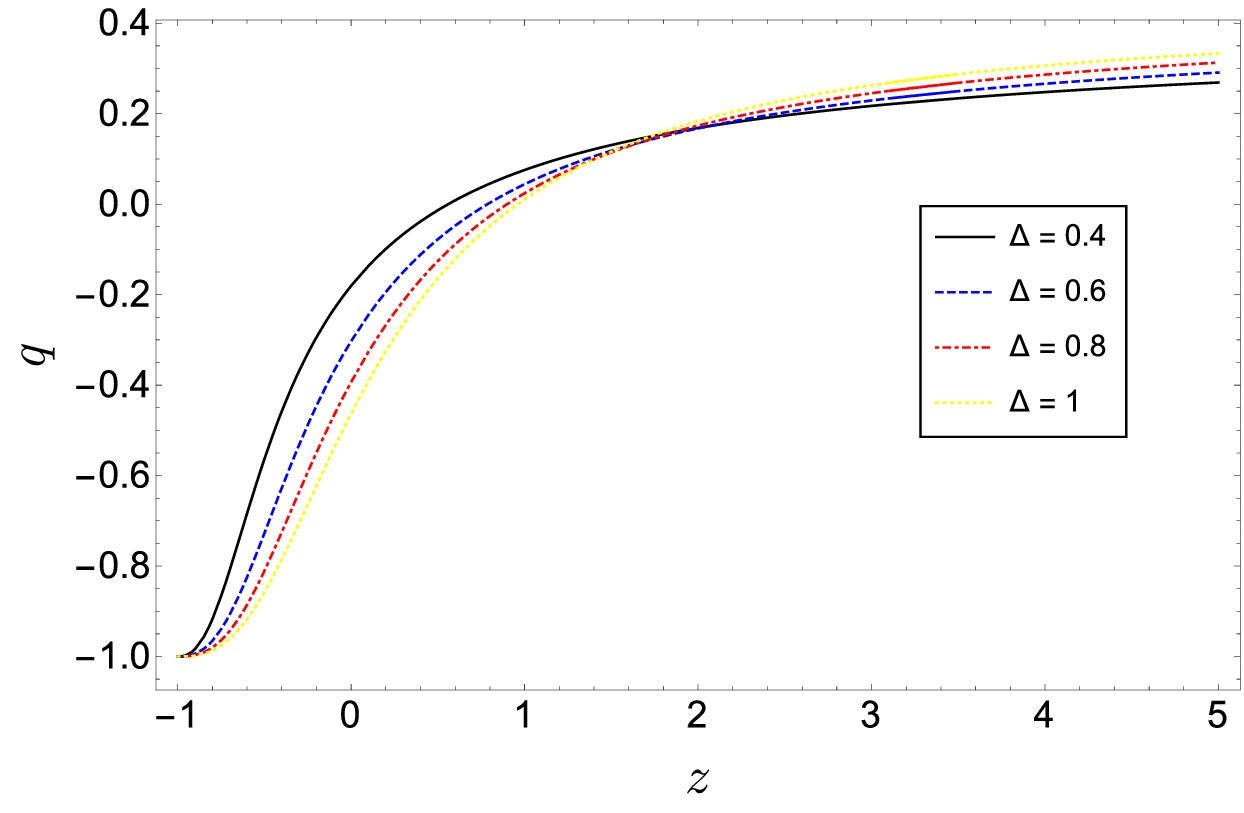}
 \caption{The evolution of $q$ versus $z$ for different values
of $\Delta$. We set $b_1^2=0.05$, $b_2^2=0.04$ and $\Omega_D^0=0.73$ as initial condition.}
 \label{Fig6}
    \end{minipage}
\end{figure}

%

Now, in order to explore the expansion history of the Universe,  
we consider the deceleration parameter 
\be
q=-\frac{\ddot a}{aH^2}=-1-\frac{\dot H}{H^2}\,.
\ee
From this relation, it follows that $q>0$ corresponds to
a decelerated expansion (i.e. $\ddot a<0$), while 
$q<0$ indicates an accelerated phase ($\ddot a>0$). 
For the present model, one can show that~\cite{Mamon:2020spa}
\be
q=\frac{-\left(1+\Delta\right)\Omega_D+1-3b_1^2+3\left(b_1^2-b_2^2\right)\Omega_D}{2\left[1-\left(1-\frac{\Delta}{2}\right)\Omega_D\right]}\,.
\ee
The evolution of $q$ versus $z$ is plotted in Fig.~\ref{Fig5} and
Fig.~\ref{Fig6} for different values of $b_1^2,b_2^2$ and $\Delta$, respectively.  As we can see, the present model 
predicts the sequence of an early matter dominated era with a decelerated expansion (at high redshift) and
a late-time DE dominated era with an accelerated 
phase (at low redshift), in contrast to the standard Holographic Dark Energy model. The transition redshift $z_t$ such that $q(z_t)=0$, lies within the intervals $0.56\le z_t\le0.83$ (Fig.~\ref{Fig5})
and $0.53\le z_t\le0.96$ (Fig.~\ref{Fig6})
for the considered values of the model parameters, consistently
with the recent result $0.5<z_t<1$ of~\cite{Rapetti,Farooq,MamonEPJC,MamonEPJC2}.
We can also estimate the current value of the deceleration parameter
as $-0.26\le q_0\le-0.20$ (Fig.~\ref{Fig5}) and
$-0.46\le q_0\le-0.17$ (Fig.~\ref{Fig6}),
to be compared with the observational result $q_0=-0.64\pm0.22$ found in~\cite{WuHu} from Union2 SNIa data.

\begin{figure}
    \centering
    \begin{minipage}{0.45\textwidth}
        \centering
\includegraphics[width=8.5cm]{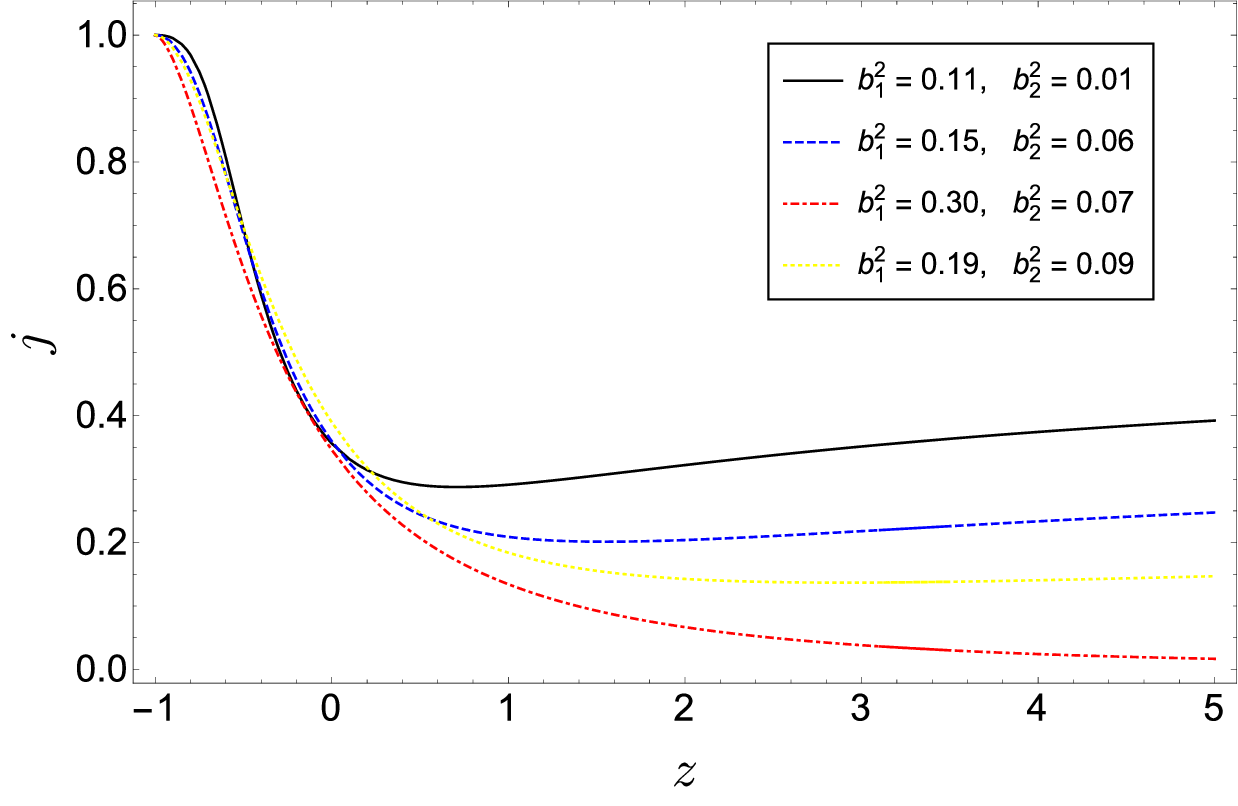}
 \caption{The evolution of $j$ versus $z$ for different values
of $b_1^2$ and $b_2^2$. We set $\Delta=0.5$ and $\Omega_D^0=0.73$ as initial condition.}
 \label{Fig7}
     \end{minipage}\hfill
    \begin{minipage}{0.45\textwidth}
        \centering
\hspace{-0.5cm}\includegraphics[width=8.5cm]{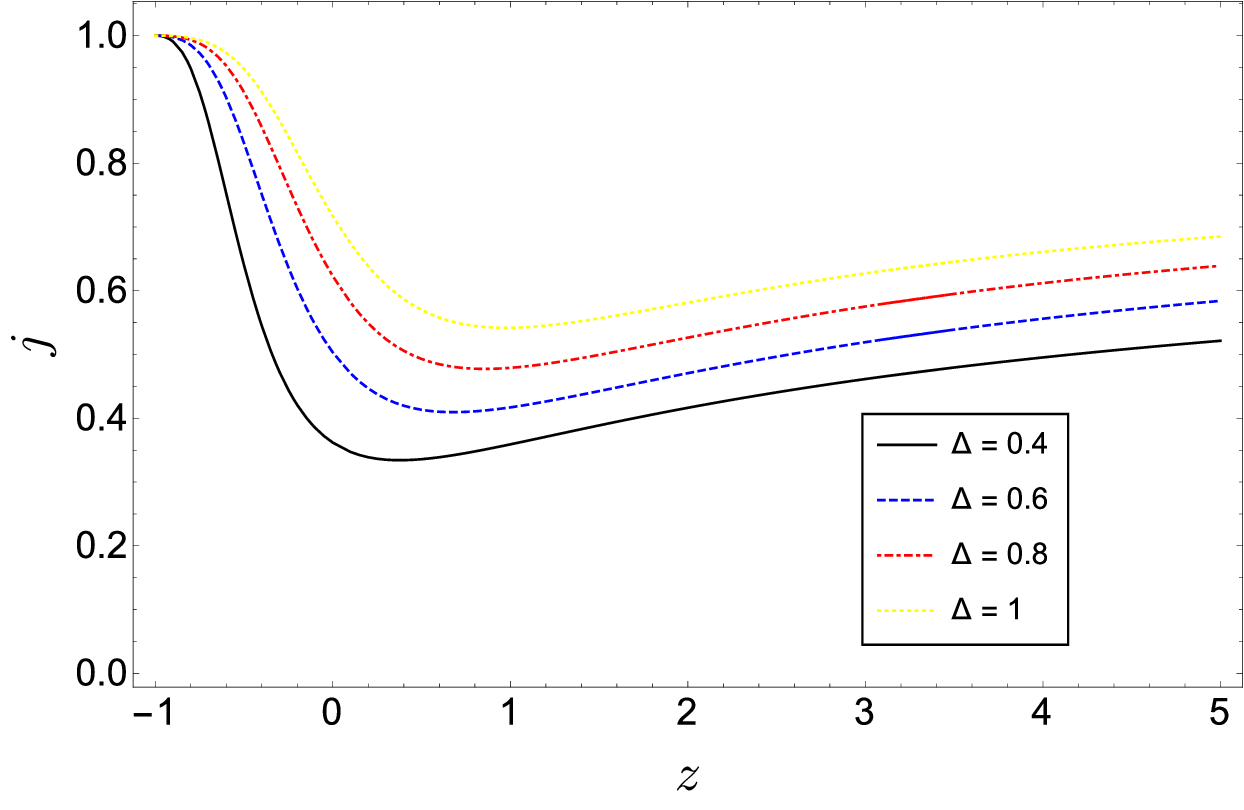}
 \caption{The evolution of $j$ versus $z$ for different values
of $\Delta$. We set $b_1^2=0.05$, $b_2^2=0.04$ and $\Omega_D^0=0.73$ as initial condition.}
 \label{Fig8}
    \end{minipage}
\end{figure}

In Fig.~\ref{Fig7} and Fig.~\ref{Fig8} we 
present the evolution of the jerk parameter $j$. This 
is a dimensionless third derivative of the scale factor
respect to the cosmic time, i.e.~\cite{Blandford:2004ah,Rapetti}
\begin{eqnarray}
\nonumber
j&=&\frac{1}{aH^3}\frac{d^3a}{dt^3}=q\left(2q+1\right)+\left(1+z\right)\frac{dq}{dz}\\[2mm]
\nonumber
&=&\frac{1}{2}\,\Bigg\{
-\frac{9\Delta\Omega_D\left(\Delta-2\right)\left[1-b_1^2+\left(b_1^2-b_2^2-1\right)\Omega_D\right]\left\{2b_1^2+\left(2b_2^2-2b_1^2+\Delta\right)\Omega_D\right\}}
{\left[2+\left(\Delta-2\right)\Omega_D\right]^3}
+\frac{3}{2+\left(\Delta-2\right)\Omega_D}\\[2mm]
\nonumber
&&\times\,\Big\{-\frac{6\Delta\left(b_1^2-b_2^2\right)\left[1-b_1^2+\left(b_1^2-b_2^2-1\right)\Omega_D\right]\Omega_D+3\Delta^2\left[1-b_1^2+\left(b_1^2-b_2^2-1\right)\Omega_D\right]
\Omega_D}
{2+\left(\Delta-2\right)\Omega_D}
\Big\}\\[2mm]
\nonumber
&&+\,\Big\{1+\frac{3\big\{-\left(2b_2^2+\Delta\right)\Omega_D+2b_1^2\left[-1+\Omega_D\right]\big\}}
{2+\left(\Delta-2\right)\Omega_D}\Big\}
\Big\{2+\frac{3\big\{-\left(2b_2^2+\Delta\right)\Omega_D+2b_1^2\left[-1+\Omega_D\right]\big\}}
{2+\left(\Delta-2\right)\Omega_D}
\Big\}
\Bigg\}\,.
\end{eqnarray}
This parameter provides us with the simplest approach to 
search for departures of the present model from $\Lambda$CDM, which
is characterized by $j=1$. 
From Fig.~\ref{Fig7} and Fig.~\ref{Fig8}, we see that $j>0$ for any
redshift and approaches unity in the far future (i.e. $z\rightarrow-1$). Therefore, in this limit our model resembles 
$\Lambda$CDM~\cite{Blandford:2004ah,Rapetti}.

%

In order to understand the classical stability of BHDE model,
let us now focus on the square of sound speed (notice that
neither this quantity nor the jerk parameter have been considered in the analysis of~\cite{Mamon:2020spa}). This is given by
\be
\label{vsq}
v_s^2=\frac{dp_D}{d\rho_D}=\omega_D+\dot\omega_D \frac{\rho_D}{\dot\rho_D}\,.
\ee
By resorting to Eqs.~\eqref{Q1},~\eqref{omegaD} and~\eqref{Omp}
and observing that
\be
\dot\omega_D\equiv\frac{d\omega_D}{dt}=\frac{d\omega_D}{d\Omega_D}\,\Omega'_D\,H\,,
\ee
we get
\be
v_s^2= \frac{b_1^2}{\frac{1}{2}\left(\Delta-2\right)\Omega_D\left[1+\frac{\Omega_D}{2}\left(\Delta-2\right)\right]^2}\,+\,\frac{-\frac{\Delta}{2}-b_2^2+b_1^2\left(2+\frac{\Delta}{2}\right)+\Omega_D\left(\frac{\Delta}{2}-b_1^2+b_2^2\right)}{\left[1+\frac{\Omega_D}{2}\left(\Delta-2\right)\right]^2}\,,
\ee
where we have expressed $H$ in terms of $\Omega_D$ by using the inverse of 
Eq.~\eqref{Omd}. 

\begin{figure}
    \centering
    \begin{minipage}{0.45\textwidth}
        \centering
\includegraphics[width=8.5cm]{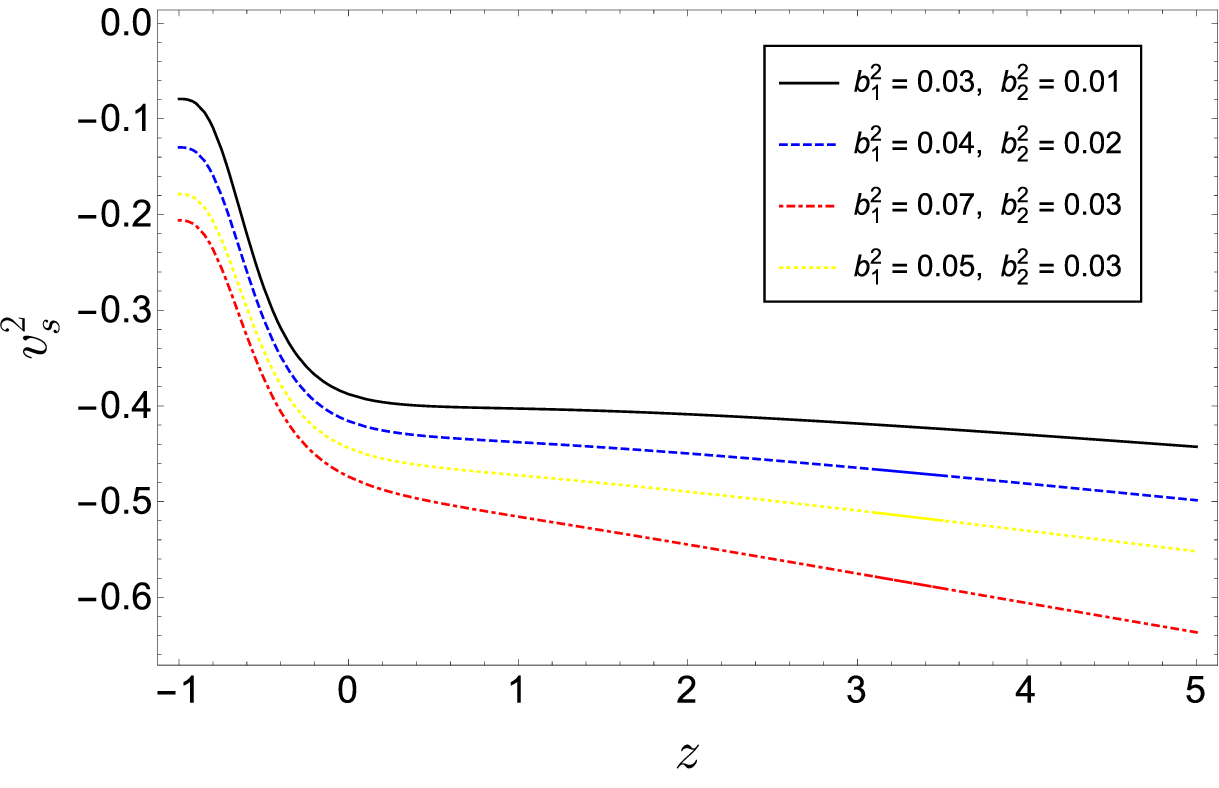}
 \caption{The evolution of $v_s^2$ versus $z$ for different values
of $b_1^2$ and $b_2^2$. We set $\Delta=0.5$ and $\Omega_D^0=0.73$ as initial condition.}
 \label{Fig9}
     \end{minipage}\hfill
    \begin{minipage}{0.45\textwidth}
        \centering
\hspace{-0.5cm}\includegraphics[width=8.5cm]{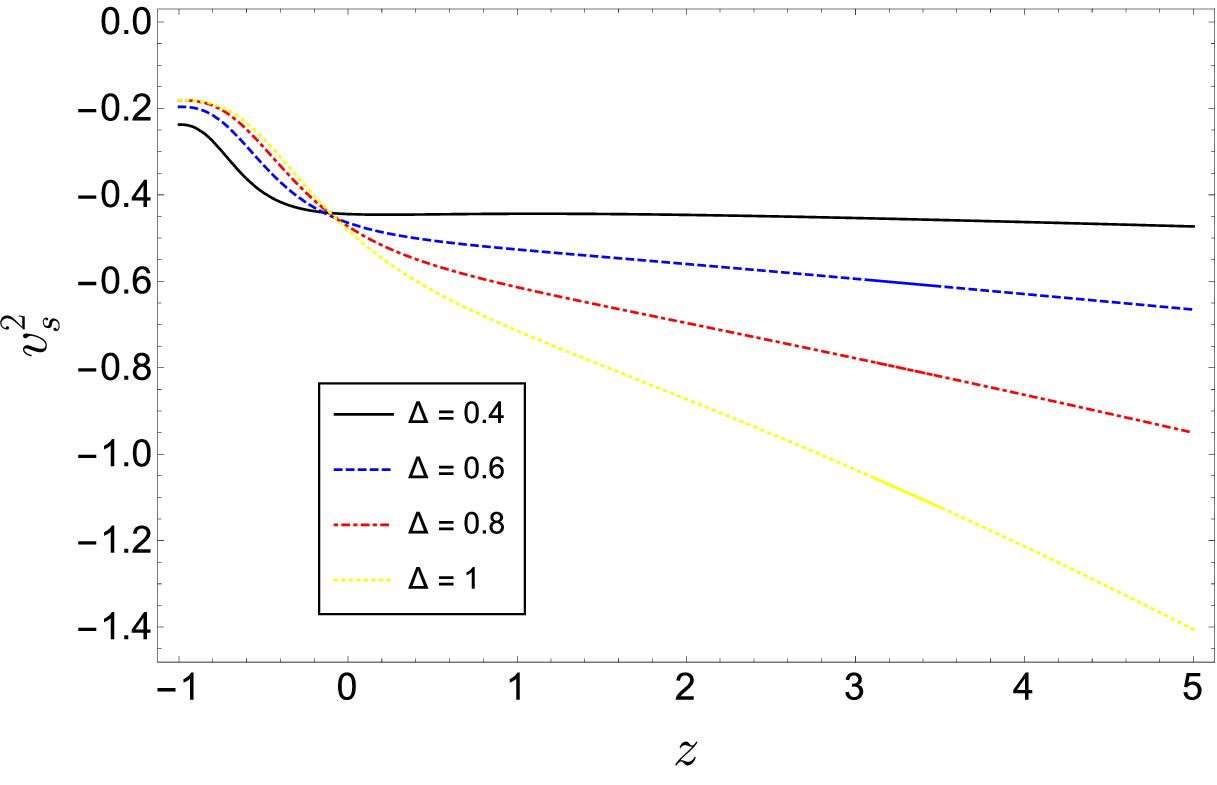}
\caption{The evolution of $v_s^2$ versus $z$ for different values
of $\Delta$. We set $b_1^2=0.05$, $b_2^2=0.04$ and $\Omega_D^0=0.73$ as initial condition.}
 \label{Fig10}
    \end{minipage}
\end{figure}

The evolutionary trajectories
of $v_s^2$ for different values of $b_1^2$, $b_2^2$ and $\Delta$
have been plotted in Fig.~\ref{Fig9} and Fig.~\ref{Fig10}, 
respectively. It emerges that
$v_s^2<0$ for all the considered combinations of the interactions terms
and Barrow parameters, 
which means that the present model is classically unstable against small perturbations.

%

\section{Generalized interacting BHDE model including radiation fluid}
\label{Rad}
Let us now extend the above analysis to the more realistic case where also a radiation fluid is present in the Universe. 
Denoting by $\rho_r$ the energy density of the radiation, 
Eq.~\eqref{F1} must be generalized to
\be
\label{H2bis}
H^2=\frac{1}{3m_p^2}\left(\rho_m+\rho_D+\rho_r\right).\\[2mm]
\ee
Assuming the radiation to be decoupled from BHDE
and DM, then the conservation equation 
for such component 
takes the form
\be
\label{drho}
\dot\rho_r+4H\rho_r=0\,.
\ee
We also introduce the fractional energy density
\be
\label{frOmr}
\Omega_r=\frac{\rho_r}{\rho_c}=\Omega_{r0}\left(1+z\right)^4\,,
\ee
which allows us to rewrite Eq.~\eqref{H2bis} as
\be
\label{newOm1}
\Omega_m+\Omega_D+\Omega_r=1\,.
\ee
If we now differentiate Eq.~\eqref{H2bis} and
then insert Eqs.~\eqref{Q1},~\eqref{drho}
and~\eqref{newOm1}, we are led to
\be
\label{dotHbis}
\frac{\dot H}{H^2}=\frac{1}{2}\left[\Omega_m+\Omega_D\left(1-3\omega_D\right)\right]-2\,,
\ee
which coincides with Eq.~\eqref{dotH} for $\Omega_r=0$, 
as expected. 

In turn, the EoS parameter~\eqref{omegaD} becomes
\be
\label{omegaDbis}
\omega_D=\frac{-\frac{\Delta}{2}+3\left(1-\frac{\Delta}{2}\right)\Omega_r}
{\left[1+\frac{\Omega_D}{2}\left(\Delta-2\right)\right]}\,+\,\frac{\left(b_1^2-b_2^2\right)\Omega_D+b_1^2\left(\Omega_r-1\right)}{\Omega_D\left[1+\frac{\Omega_D}{2}\left(\Delta-2\right)\right]}\,, 
\ee
which gives the following generalized differential equation for $\Omega_D$
\be
\label{Ompbis}
\Omega'_D=\frac{3\Delta\Omega_D}{2}\left[\frac{4-3\Omega_m-4\Omega_D-b_1^2\Omega_m-b_2^2\Omega_D}{1-\left(1-\frac{\Delta}{2}\right)\Omega_D}\right].
\ee
By use of Eq.~\eqref{newOm1}, it is easy to show that this expression matches with Eq.~\eqref{Omp} in the absence
of radiation fluid. 

\begin{figure}
    \centering
    \begin{minipage}{0.45\textwidth}
        \centering
\includegraphics[width=8.5cm]{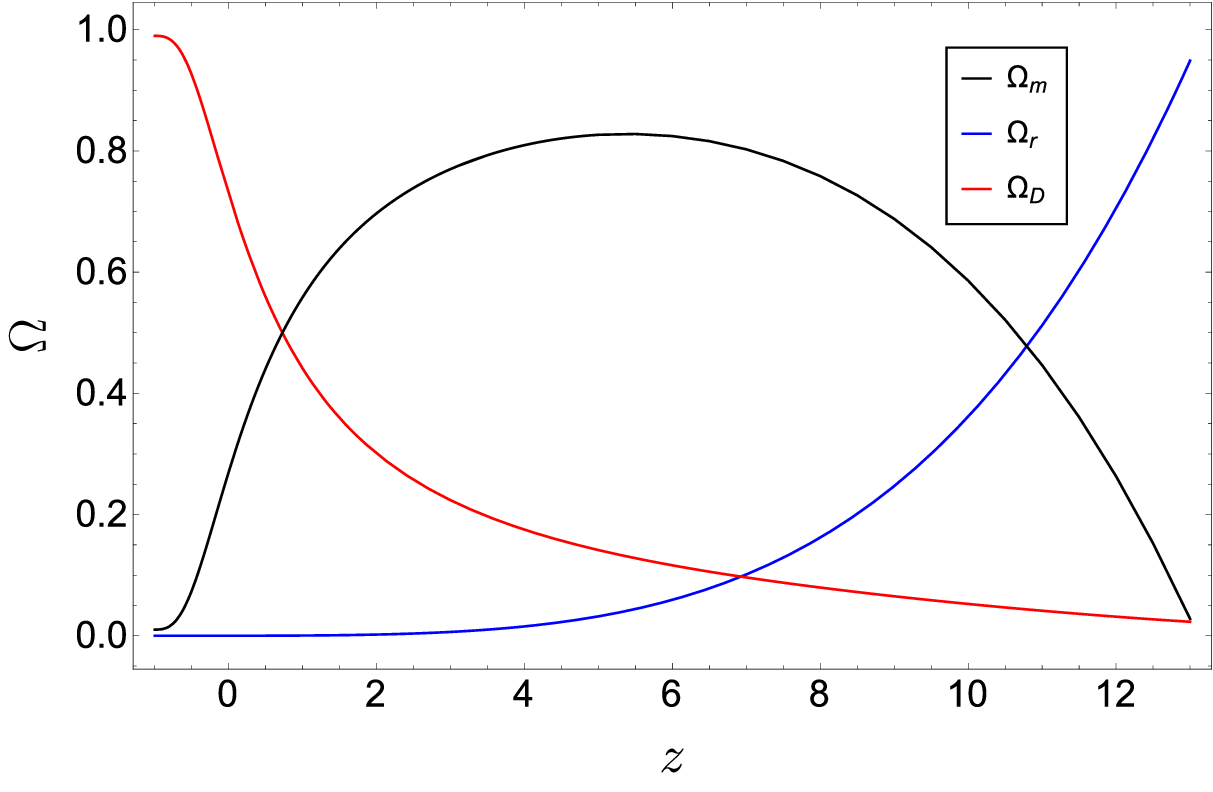}
\caption{The evolution of $\Omega_D$, $\Omega_m$ and $\Omega_r$ versus $z$. We set $\Delta=0.8$, $b_1^2=0.03$, $b_2^2=0.01$ and $\Omega_D^0=0.73$, $\Omega_r^0=2.47\times10^{-5}$ as initial conditions.}
 \label{Fig11}
      \end{minipage}\hfill
    \begin{minipage}{0.45\textwidth}
        \centering
\includegraphics[width=8.5cm]{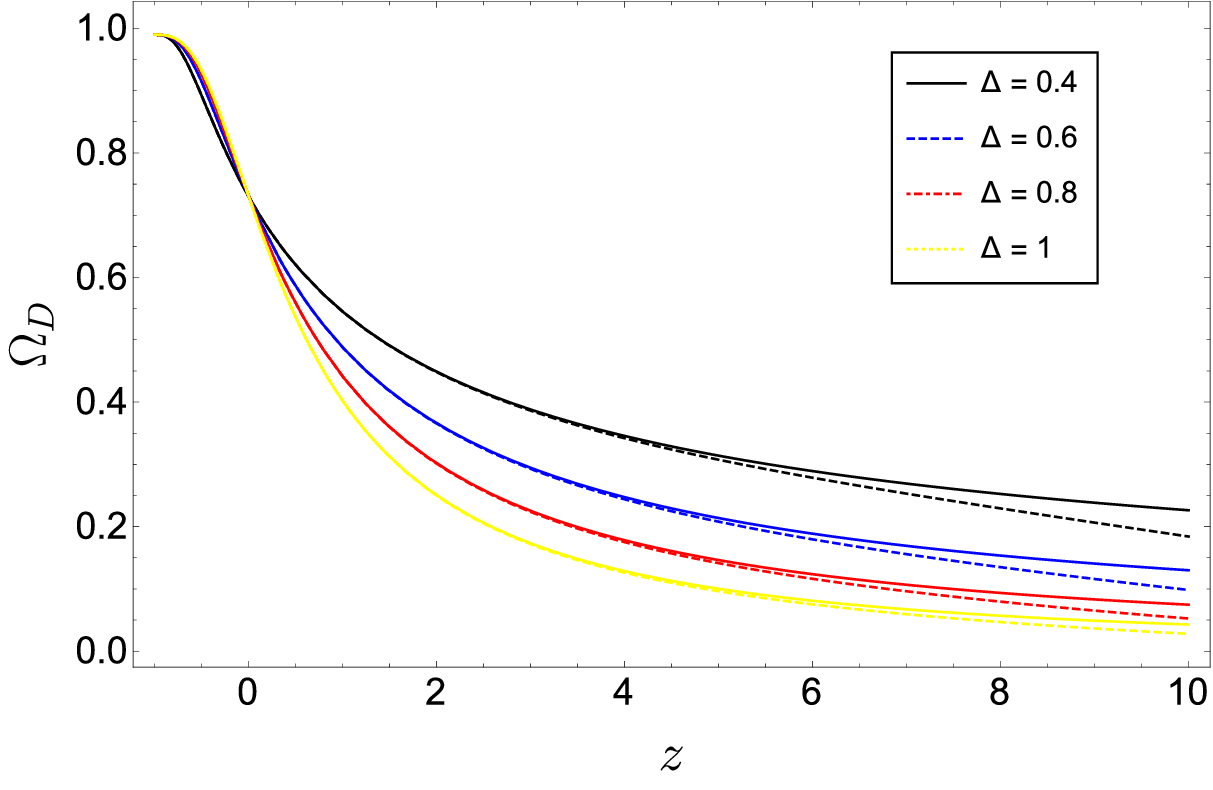}
\caption{The evolution of $\Omega_D$ versus $z$ for different values
of $\Delta$. We set $b_1^2=0.03$, $b_2^2=0.01$ and $\Omega_D^0=0.73$, $\Omega_r^0=2.47\times10^{-5}$ as initial conditions. The dashed (solid) lines represent the evolution when radiation is present (absent). }
 \label{Fig12}
    \end{minipage}
\end{figure}

As done in Sec.~\ref{InterBHDE}, we can solve Eq.~\eqref{Ompbis}
numerically (see also Appendix). The behavior of the fractional energy density parameters
$\Omega_D$, $\Omega_m$ and $\Omega_r$ is plotted in Fig.~\ref{Fig11}
for given values of $b_1^2$, $b_2^2$ and $\Delta$. It is evident
that our model well describes the usual thermal history
of the Universe, evolving from an initial radiation dominated era 
to a matter dominated one. Finally, the Universe
enters the DE dominated epoch at $0.52\le z_t\le0.82$
(Fig.~\ref{Fig15}) and 
$0.52\le z_t\le0.97$ (Fig.~\ref{Fig16}) for the considered
values of model parameters. These intervals
are again consistent with the observational predictions of~\cite{Rapetti,Farooq,MamonEPJC,MamonEPJC2}. 

On the other hand, in Fig.~\ref{Fig12} we plot
the evolution of $\Omega_D$ in the presence (dashed lines)
and absence (solid lines) of radiation fluid for different 
values of $\Delta$. As expected,  the 
discrepancy between the two sets of curves
becomes more evident at higher redshift, where
the effects of radiation are more relevant (see Fig.~\ref{Fig11}). 
Interestingly enough, we also notice that 
the larger the value of $\Delta$, the lower the magnitude of the difference between the two sets of curves. This is due
to the fact that, for higher $\Delta$, the contribution of BHDE density
increases at the expense of the radiation
term, whose effects then become increasingly negligible. 
A similar result has been exhibited in~\cite{Mamon:2020wnh} for the case of the interacting Tsallis Holographic Dark Energy.

%

\begin{figure}[t]
    \centering
    \begin{minipage}{0.45\textwidth}
        \centering
\includegraphics[width=8.5cm]{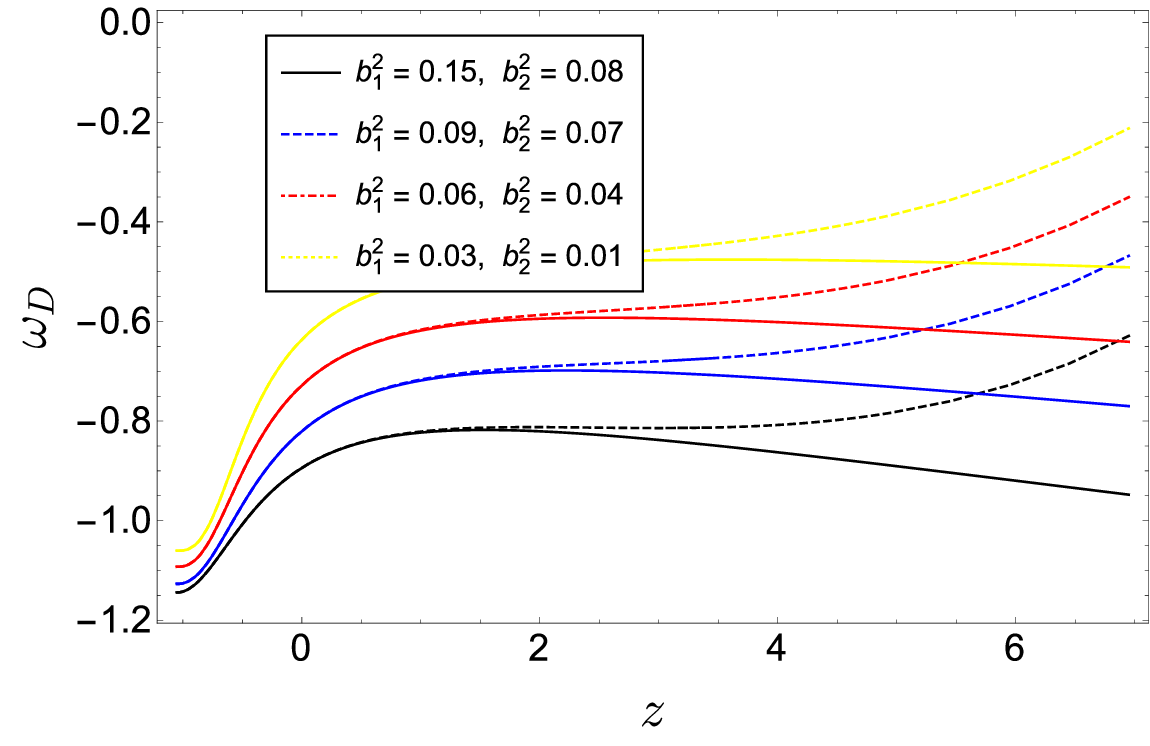}
\caption{The evolution of $\omega_D$ versus $z$ for different values
of $b_1^2$ and $b_2^2$. We set $\Delta=0.5$ and $\Omega_D^0=0.73$, $\Omega_r^0=2.47\times10^{-5}$ as initial conditions. The dashed (solid) lines represent the evolution when radiation is present (absent). }
\label{Fig13}
     \end{minipage}\hfill
    \begin{minipage}{0.45\textwidth}
        \centering
\hspace{-0.5cm}\includegraphics[width=8.5cm]{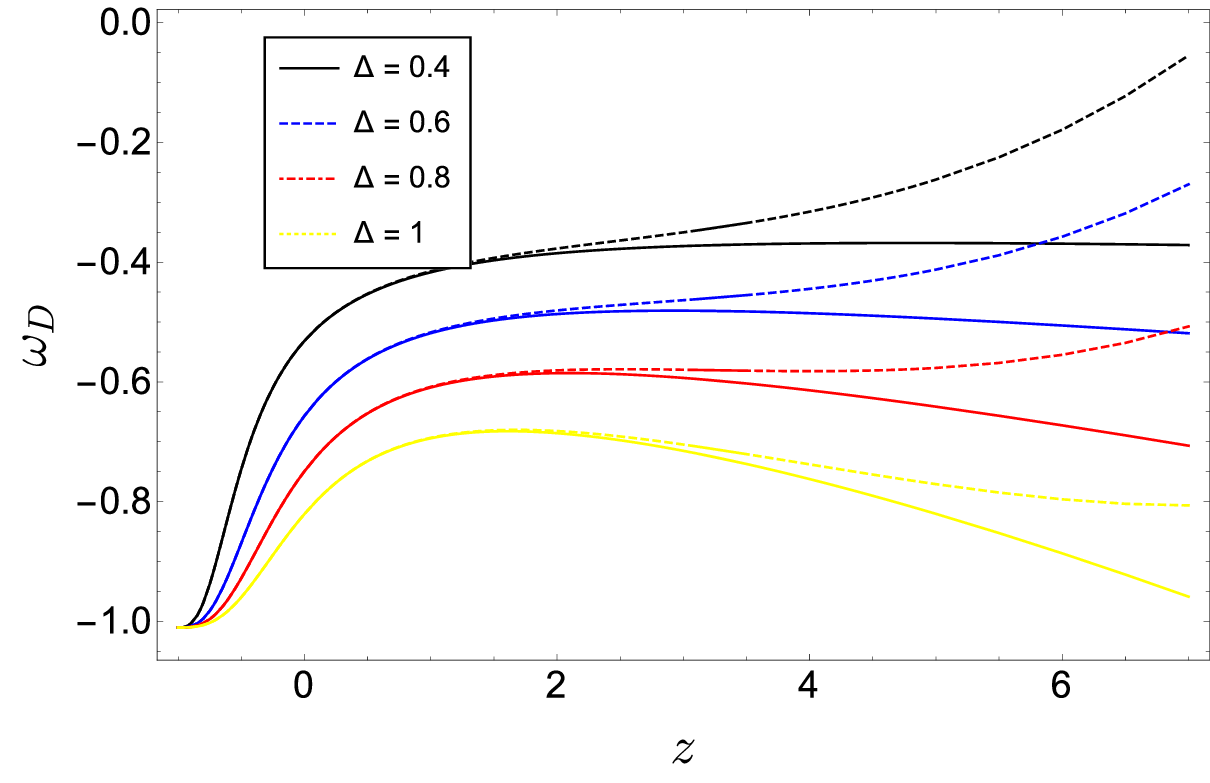}
\caption{The evolution of $\omega_D$ versus $z$ for different values
of $\Delta$. We set $b_1^2=0.03$, $b_2^2=0.01$ and $\Omega_D^0=0.73$, $\Omega_r^0=2.47\times10^{-5}$ as initial conditions. The dashed (solid) lines represent the evolution when radiation is present (absent). }
\label{Fig14}
    \end{minipage}
\end{figure}

\begin{figure}[H]
    \centering
    \begin{minipage}{0.45\textwidth}
        \centering
\includegraphics[width=8.5cm]{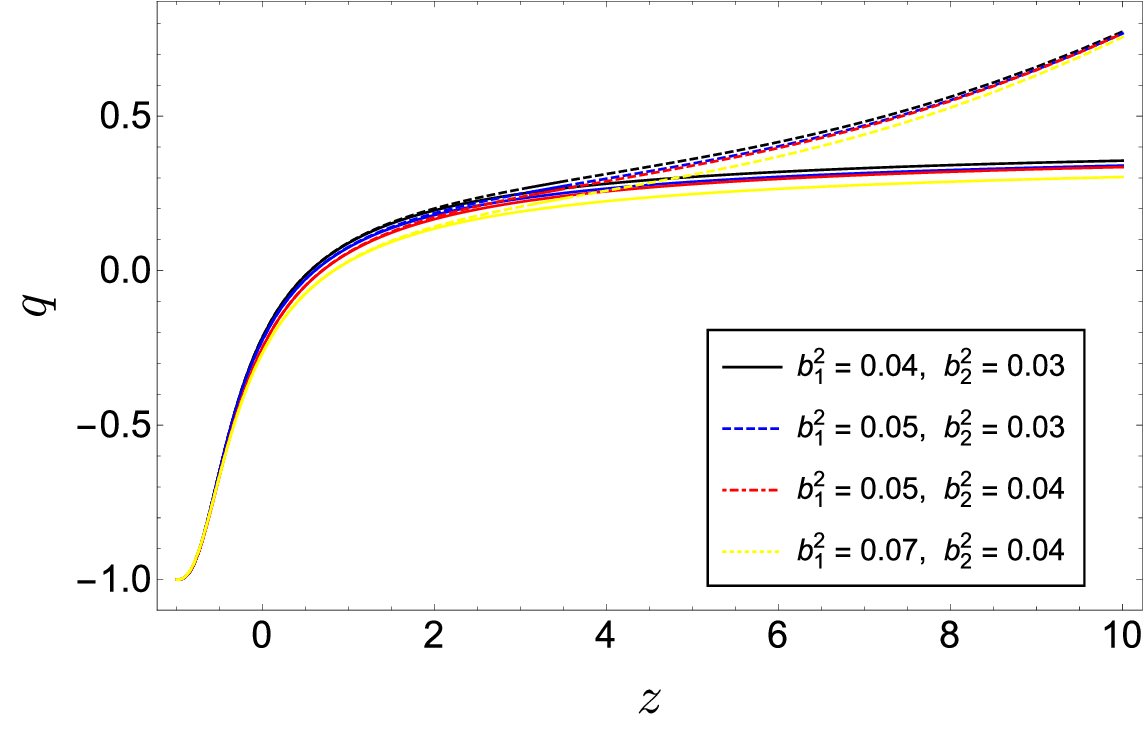}
\caption{The evolution of $q$ versus $z$ for different values
of $b_1^2$ and $b_2^2$. We set $\Delta=0.5$ and $\Omega_D^0=0.73$, $\Omega_r^0=2.47\times10^{-5}$ as initial conditions. The dashed (solid) lines represent the evolution when radiation is present (absent). }
\label{Fig15}
     \end{minipage}\hfill
    \begin{minipage}{0.45\textwidth}
        \centering
\hspace{-0.5cm}\includegraphics[width=8.5cm]{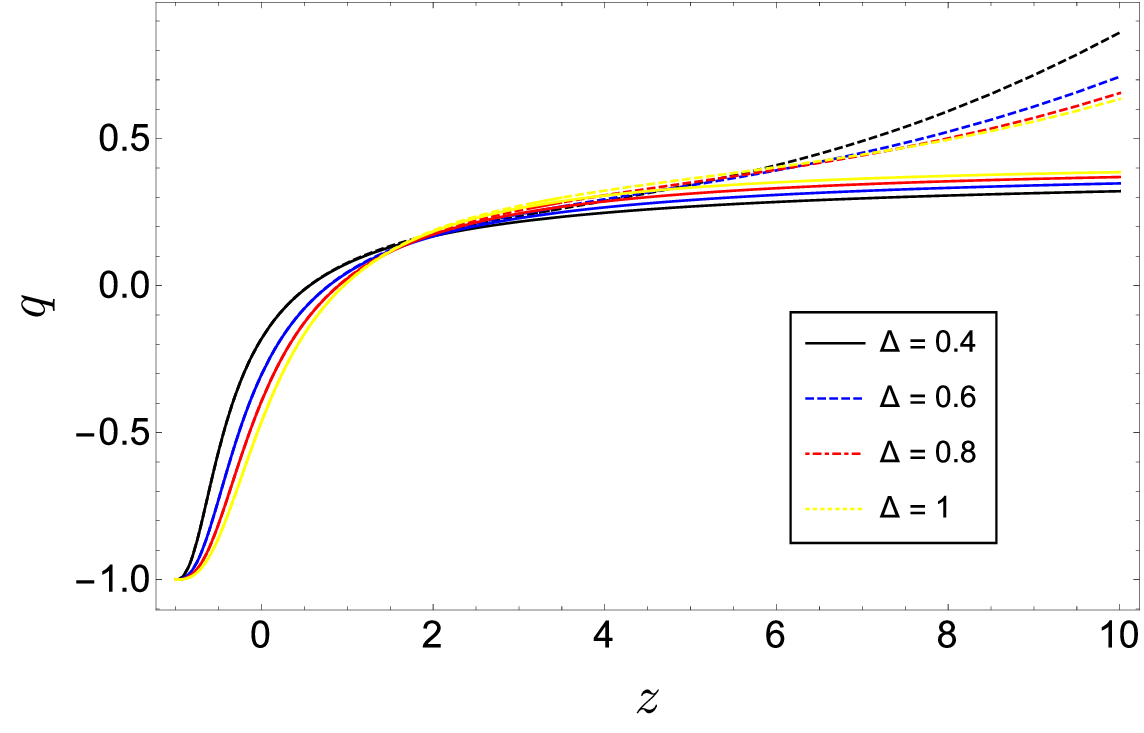}
\caption{The evolution of $q$ versus $z$ for different values
of $\Delta$. We set $b_1^2=0.05$, $b_2^2=0.04$ and $\Omega_D^0=0.73$, $\Omega_r^0=2.47\times10^{-5}$ as initial conditions. The dashed (solid) lines represent the evolution when radiation is present (absent). }
\label{Fig16}
    \end{minipage}
\end{figure}

The behavior of the BHDE EoS parameter~\eqref{omegaDbis} is
presented in Fig.~\ref{Fig13} for different values of the interaction terms $b_1^2$ and $b_2^2$. Again, for lower redshift the dashed and solid lines nearly coincide, which explains why the prediction 
for the present value of  $\omega_D$ is the same as that
found in the previous Section.\footnote{A similar reasoning applies
to the other parameters calculated below.}
In particular, in the interval where the
transition from the decelerated to accelerated phase occurs, i.e. $0.5<z_t<1$, BHDE 
lies in the quintessence dominated phase, while it evolves 
toward a phantom-like behavior for $z\rightarrow-1$. 
However, as $z$ increases, the two sets of curves depart
from each other, in line with the previous discussion.  
The same profile is exhibited for different values of $\Delta$
(see Fig.~\ref{Fig14}).

%

\begin{figure}[t]
\begin{center}
\includegraphics[width=8.5cm]{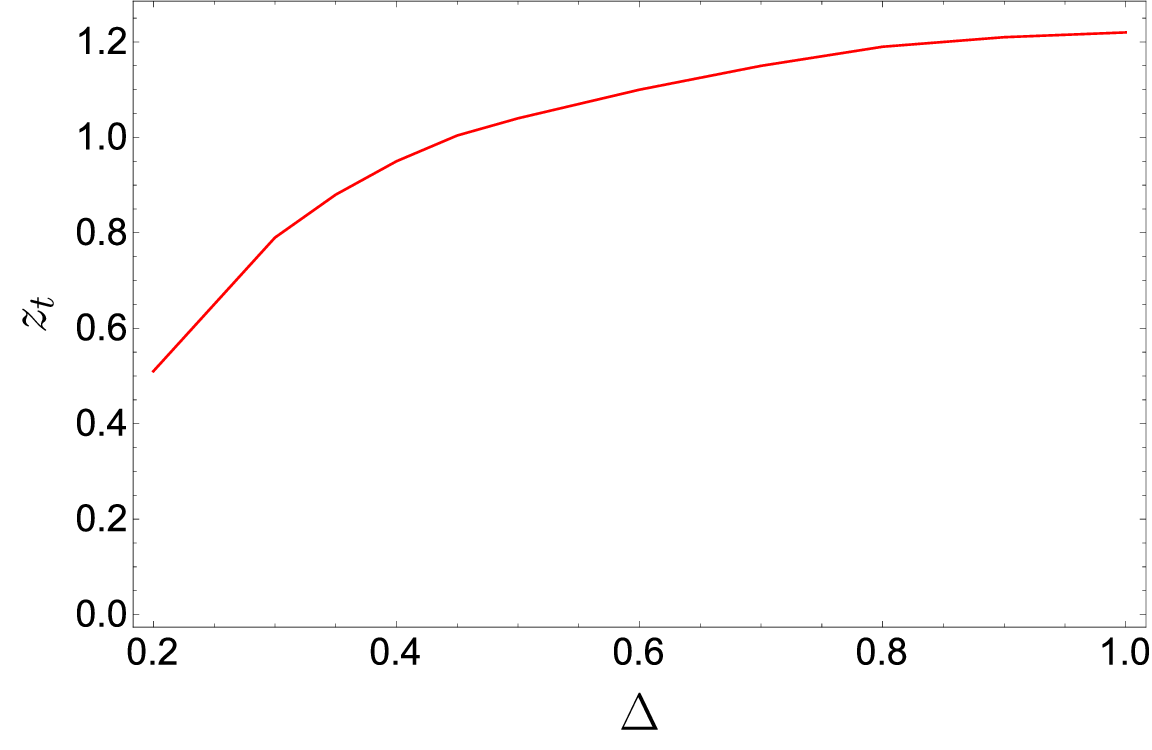}
\caption{Transition redshift $z_t$ versus $\Delta$. We set $b_1^2=0.07$, $b_2^2=0.06$ and $\Omega_D^0=0.73$, $\Omega_r^0=2.47\times10^{-5}$ as initial conditions.}
\label{Fig17}
\end{center}
\end{figure}

In Fig.~\ref{Fig15} and Fig.~\ref{Fig16} we plot the deceleration parameter 
\be
\label{decbis}
q=\frac{1}{2}\left\{1+\Omega_r^0\left(1+z\right)^4
\,+\,\frac{3\left[-\left(2b_2^2+\Delta\right)\Omega_D-3\Omega_r^0\left(\Delta-2\right)\left(1+z\right)^4\Omega_D\right]+6b_1^2\left[-1+\Omega_D+\Omega_{r}^0\left(1+z\right)^4\right]}
{2+\left(\Delta-2\right)\Omega_D}
\right\}
\ee
versus $z$ for different values of $b_1^2,b_2^2$ and $\Delta$, 
respectively. We see that also in the present model
the Universe evolves from an early decelerated regime
(for $z>z_t$) to a late accelerated phase (for $z<z_t$).
The value of $z_t$ strongly depends on Barrow parameter
$\Delta$, with $z_t$ increasing for increasing $\Delta$
(see Fig.~\ref{Fig17}). However, for $\Delta\gtrsim 0.55$ 
and the selected values of $b_1^2$ and $b_2^2$,
we find that $z_t>1$, 
in contrast with the results of~\cite{Rapetti,Farooq,MamonEPJC,MamonEPJC2}. Thus, 
large values of $\Delta$ turn out to be disfavored by our model.
This is consistent with the stringent upper bounds 
on $\Delta$ obtained through different cosmological analysis 
in~\cite{Anagnostopoulos:2020ctz,Leon:2021wyx,Barrow:2020kug,Jusufi:2021fek,Dabrowski:2020atl,Saridakis:2020cqq,Luciano:2022pzg,Vagnozzi:2022moj}.

%
%
%
%

\begin{figure}
    \centering
    \begin{minipage}{0.45\textwidth}
        \centering
\includegraphics[width=8.5cm]{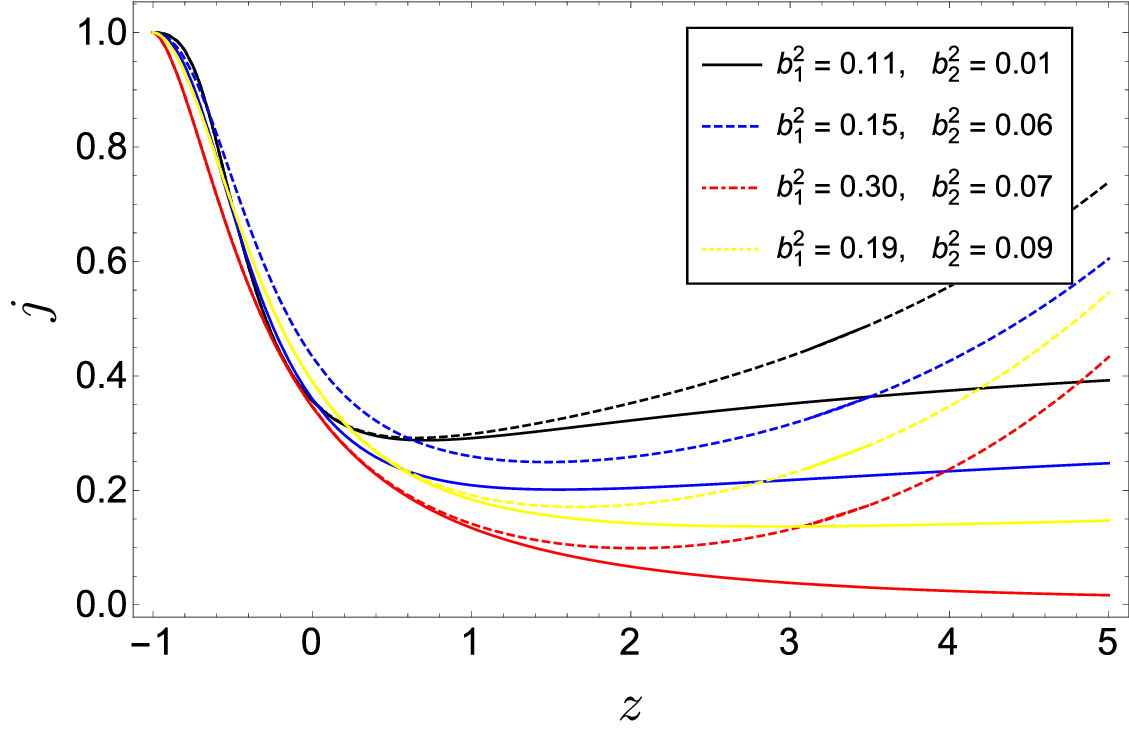}
\caption{The evolution of $j$ versus $z$ for different values
of $b_1^2$ and $b_2^2$. We set $\Delta=0.5$ and $\Omega_D^0=0.73$, $\Omega_r^0=2.47\times10^{-5}$ as initial conditions. The dashed (solid) lines represent the evolution when radiation is present (absent). }
\label{Fig18}
     \end{minipage}\hfill
    \begin{minipage}{0.45\textwidth}
        \centering
\hspace{-0.5cm}\includegraphics[width=8.5cm]{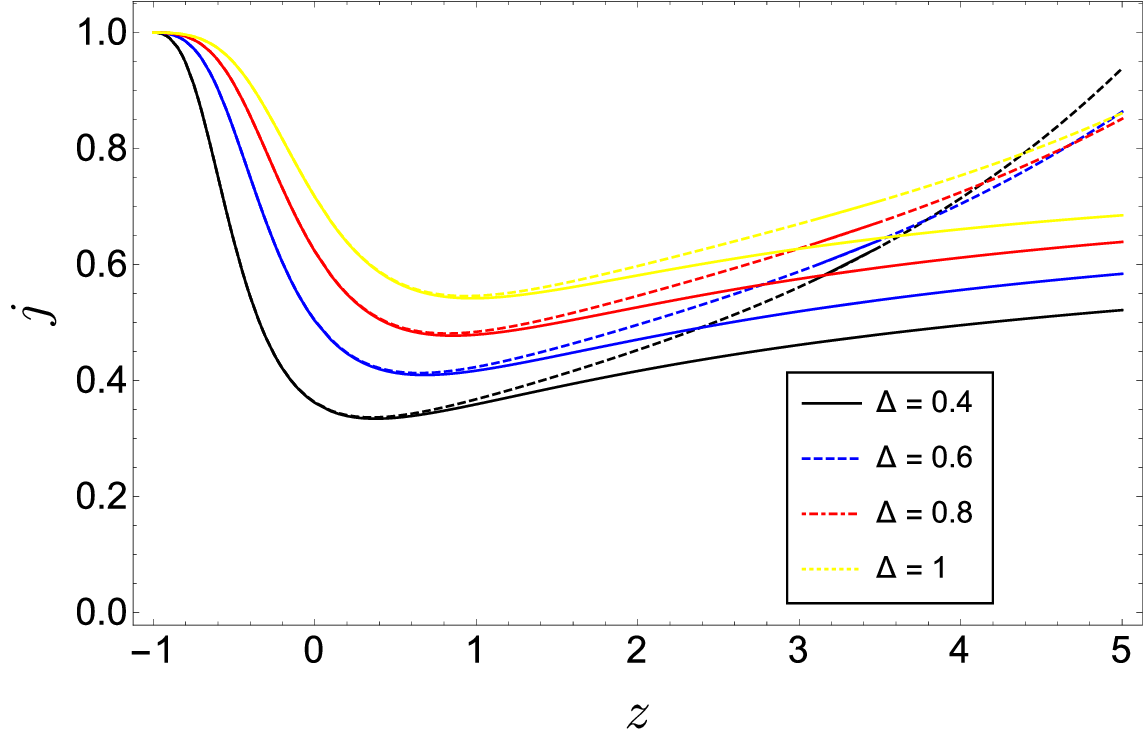}
\caption{The evolution of $j$ versus $z$ for different values
of $\Delta$. We set $b_1^2=0.05$, $b_2^2=0.04$ and $\Omega_D^0=0.73$, $\Omega_r^0=2.47\times10^{-5}$ as initial conditions. The dashed (solid) lines represent the evolution when radiation is present (absent). }
\label{Fig19}
    \end{minipage}
\end{figure}

Figure~\ref{Fig18} and Fig.~\ref{Fig19} show the
behavior of the jerk parameter 
\begin{eqnarray}
\label{jbis}
\nonumber
j&=&\frac{1}{2}\,\Bigg\{4\Omega_r^0\left(1+z\right)^4
-\frac{9\Delta\left(\Delta-2\right)\left[1-b_1^2+\left(b_1^2-b_2^2-1\right)\Omega_D+\Omega_{r}^0\left(3+b_1^2\right)\left(1+z\right)^4\right]\Omega_D}
{\left[2+\left(\Delta-2\right)\Omega_D\right]^3}\\[2mm]
\nonumber
&&\times\left\{2b_1^2+\left(2b_2^2-2b_1^2+\Delta\right)\Omega_D-\Omega_r^0\left(1+z\right)^4\left[2b_1^2-3\left(\Delta-2\right)\Omega_D\right]\right\}
+\frac{3}{2+\left(\Delta-2\right)\Omega_D}\\[2mm]
\nonumber
&&\times\,\Big\{8b_1^2\Omega_r^0\left(1+z\right)^4-12\Omega_r^0\left(\Delta-2\right)\left(1+z\right)^4\Omega_D-\frac{6\Delta\left(b_1^2-b_2^2\right)\left[1-b_1^2+\left(b_1^2-b_2^2-1\right)\Omega_D+\Omega_r^0\left(3+b_1^2\right)\left(1+z\right)^4\right]\Omega_D}
{2+\left(\Delta-2\right)\Omega_D}\\[2mm]
\nonumber
&&+\,\frac{3\Delta\left[1-b_1^2+\left(b_1^2-b_2^2-1\right)\Omega_D+\Omega_r^0\left(3+b_1^2\right)\left(1+z\right)^4\right]
\left[\Delta+3\Omega_r^0\left(\Delta-2\right)\left(1+z\right)^4\right]\Omega_D}
{2+\left(\Delta-2\right)\Omega_D}
\Big\}\\[2mm]
\nonumber
&&+\,\Big\{1+\Omega_r^0\left(1+z\right)^4+\frac{3\big\{-\left(2b_2^2+\Delta\right)\Omega_D-3\Omega_r^0\left(\Delta-2\right)\left(1+z\right)^4\Omega_D+2b_1^2\left[-1+\Omega_D+\Omega_r^0\left(1+z\right)^4\right]\big\}}
{2+\left(\Delta-2\right)\Omega_D}\Big\}\\[2mm]
&&\times\,
\Big\{2+\Omega_r^0\left(1+z\right)^4+\frac{3\big\{-\left(2b_2^2+\Delta\right)\Omega_D-3\Omega_r^0\left(\Delta-2\right)\left(1+z\right)^4\Omega_D+2b_1^2\left[-1+\Omega_D+\Omega_r^0\left(1+z\right)^4\right]\big\}}
{2+\left(\Delta-2\right)\Omega_D}
\Big\}
\Bigg\}
\end{eqnarray}
versus the 
redshift $z$. As in the absence of radiation, we
find that $j>0$ and tends to unity in the far future (i.e. $z\rightarrow-1$).
Once more, the discrepancy
between the curves representing the evolution of $j$ with (dashed lines)
and without (solid lines) radiation becomes greater
for higher redshift, where the contribution 
of radiation dominates.

%
%

\begin{figure}
    \centering
    \begin{minipage}{0.45\textwidth}
        \centering
\includegraphics[width=8.5cm]{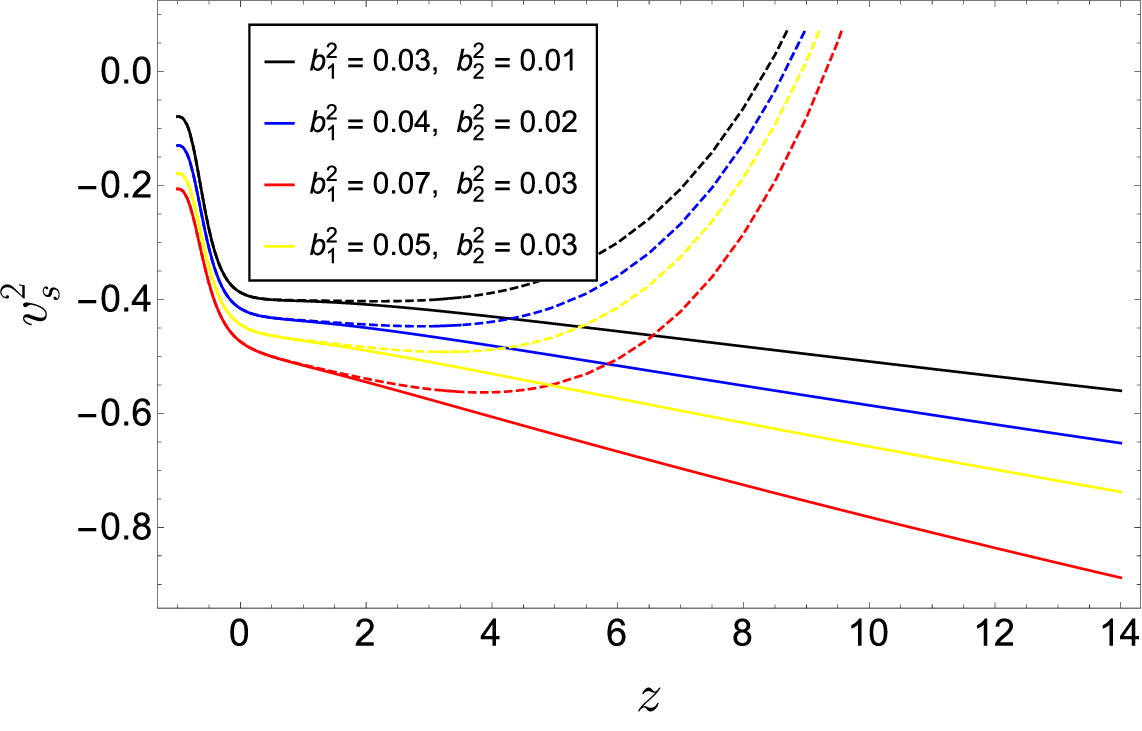}
\caption{The evolution of $v_s^2$ versus $z$ for different values
of $b_1^2$ and $b_2^2$. We set $\Delta=0.5$ and $\Omega_D^0=0.73$, $\Omega_r^0=2.47\times10^{-5}$ as initial conditions. The dashed (solid) lines represent the evolution when radiation is present (absent). }
\label{Fig20}
     \end{minipage}\hfill
    \begin{minipage}{0.45\textwidth}
        \centering
\hspace{-0.5cm}\includegraphics[width=8.5cm]{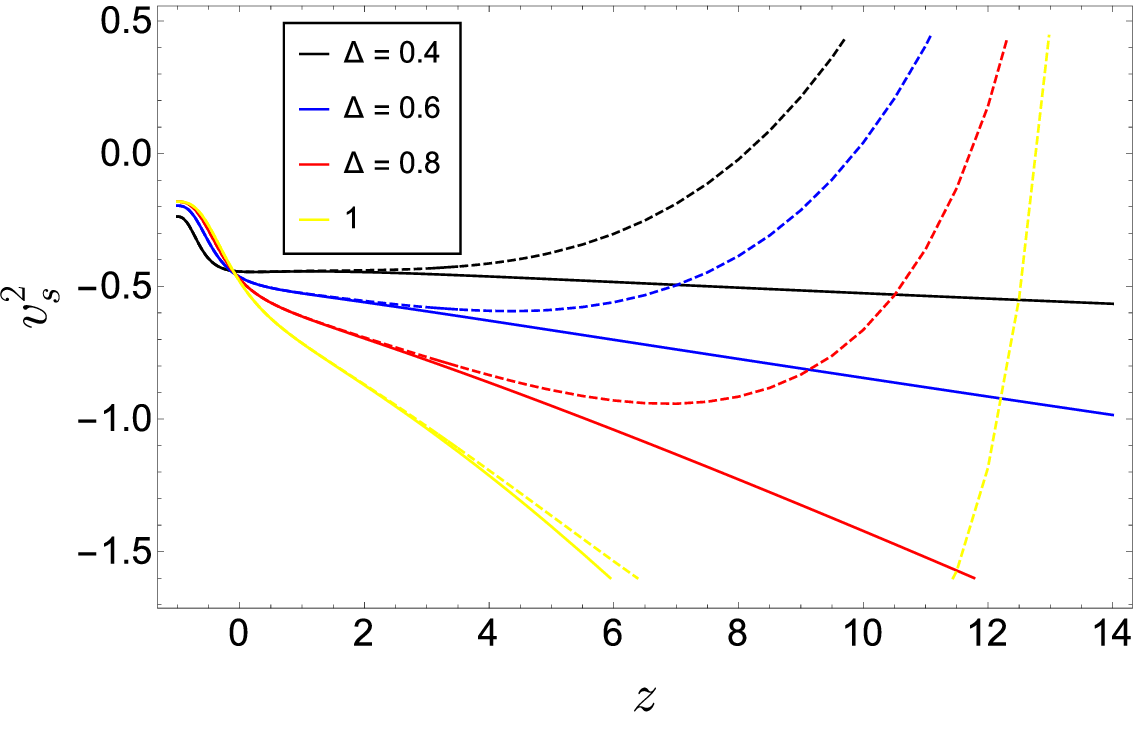}
\caption{The evolution of $v_s^2$ versus $z$ for different values
of $\Delta$. We set $b_1^2=0.05$, $b_2^2=0.04$ and $\Omega_D^0=0.73$, $\Omega_r^0=2.47\times10^{-5}$ as initial conditions. The dashed (solid) lines represent the evolution when radiation is present (absent). }
\label{Fig21}
    \end{minipage}
\end{figure}

To conclude, in Fig.~\ref{Fig20} and Fig.~\ref{Fig21}
we plot the evolution of the square of sound speed 
\begin{eqnarray}
\nonumber
v_s^2&=&\frac{1}{\Omega_D\left[2+\left(\Delta-2\right)\Omega_D\right]^2}
\Bigg\{2\left[2+\left(\Delta-2\right)\Omega_D\right]
\Big\{\left(b_1^2-b_2^2\right)\Omega_D+b_1^2\left[-1+\Omega_r^0\left(1+z\right)^4\right]
\\[2mm]
\nonumber
&&+\left[-\frac{\Delta}{2}-\frac{3}{2}\Omega_r^0\left(\Delta-2\right)\left(1+z\right)^4\right]\Omega_D\Big\}+\frac{\Delta}{\Delta-2}
\Big\{-2b_1^2\big\{\left(\Delta-2\right)\Omega_D^2+2\left[-1+\Omega_r^0\left(1+z\right)^4\right]
\\[2mm]
&&\,+2\left(\Delta-2\right)\left[-1+\Omega_r^0\left(1+z\right)^4\right]\Omega_D\big\}+\left(\Delta-2\right)\big\{2b_2^2-6\Omega_r^0\left(1+z\right)^4+\Delta\left[1+3\Omega_r^0\left(1+z\right)^4\right]
\big\}
\Omega_D^2\Big\}\Bigg\}
\end{eqnarray}
against redshift for different values of $b_1^2$, $b_2^2$ and
$\Delta$, respectively.  It is interesting to note that, 
when the effects of radiation are considered, 
the model is classically stable ($v_s^2>0$)
at early times (higher redshifts), while it 
becomes unstable at later times (lower redshifts). This feature
is peculiar to the generalized interacting BHDE
model considered in this section.

%
%

\section{Thermodynamics of generalized interacting BHDE model}
\label{Thermo}
Let us now study the thermodynamic implications of the present model.
We stress that our analysis extends that of~\cite{Mamon:2020wnh} to the case where also radiation is included in the energy budget
of the Universe.
We consider the Universe as a thermodynamic system bounded by the cosmological apparent horizon of radius~\cite{Bak}
\be
r_h=\left(H^2+\frac{k}{a^2}\right)^{-1/2}\,,
\ee
where $k$ is the spatial curvature. 
Clearly, for a spatially flat Universe, the above equation
gives $r_h=1/H$.

Denoting by $S_D$, $S_m$ and $S_r$ the entropies of BHDE, 
DM and radiation, respectively, and by $S_f=S_D+S_m+S_r$
their sum, the total entropy $S_{tot}$ of the Universe is
\be
S_{tot}=S_f+S_h\,,
\ee
where $S_h$ is the horizon entropy.
 
If we consider the Universe as an isolated system, 
then according to the generalized second law (GSL) 
of thermodynamics $S$ should obey the inequality~\cite{Mamon:2020wnh,Mamon:2020spa}
\be
\dot S_{tot}\ge0\,.
\ee
As noted in~\cite{Mamon:2020wnh}, 
this condition should hold throughout
the evolution of the Universe.

In order to establish whether the GSL is satisfied, 
let us remember that the horizon entropy $S_h$ in BHDE
is given by Eq.~\eqref{BE}, here rewritten as
\be
\label{horent}
S_h=\gamma r_h^{\Delta+2}\,,
\ee
where we have expressed the horizon area as $A=4\pi r_h^2$
and $\gamma=(4\pi/A_0)^{1+\Delta/2}>0$.
 
Furthermore, the temperature of the apparent horizon is
\be
\label{T}
T=\frac{1}{2\pi r_h}\,,
\ee
which is formally analogous to the temperature 
associated to the event horizon of a black hole with the same radius $r_h$~\cite{CaiKim}. 
At this stage, it is worth noting that such temperature
is typically assumed equal to the temperature
of the composite matter inside the horizon, otherwise
a temperature gradient might arise. 
In turn, the ensuing energy flow would imply a deformation 
of the horizon geometry and the need to resort to non-equilibrium
thermodynamics~\cite{Izquierdo,Padmanabhan:2009vy,Jamil1,Jamil2,Mimoso}.

Now, the 
first law of thermodynamics for BHDE, DM and radiation
takes the form
\be
\label{flt}
T \,dS_D = d E_D + p_D\, d V\,,\qquad T \,dS_m = d E_m\,, \qquad 
T \,dS_r = d E_r + p_r\, d V\,, 
\ee
where $V=4\pi r_h^3/3$ is the spherical volume 
occupied by the fluid inside the horizon and we have
taken into account the assumption of pressureless dark matter (i.e. $p_m=0$). Moreover, $E_D=4\pi r_h^3\rho_D/3$, $E_m=4\pi r_h^3\rho_m/3$ and $E_r=4\pi r_h^3\rho_r/3$ denote 
the  internal energies of BHDE, DM and radiation, respectively, while
$p_r=\rho_r/3$ is the radiation pressure. 

Differentiation of Eqs.~\eqref{horent} and~\eqref{flt}
respect to time gives
\be
\dot S_h=\left(\Delta+2\right)\gamma\,r_h^{\Delta+1}\dot r_h\,,
\ee
and
\be
\label{dotS}
\dot S_D=\frac{1}{T}\left(\dot E_D+4\pi\,p_D\hspace{0.5mm} r^2_h\hspace{0.5mm}\dot r_h\right),
\qquad 
\dot S_r=\frac{\dot E_m}{T}\,,\qquad
\dot S_r=\frac{1}{T}\left(\dot E_r+4\pi\,p_r\hspace{0.5mm} r^2_h\hspace{0.5mm}\dot r_h\right).
\ee
Identifying the temperature $T$ in Eq.~\eqref{dotS}
with the horizon temperature~\eqref{T}, the 
total entropy variation of the thermodynamic system bounded by the dynamical apparent horizon is
\be
\label{totentvar}
\dot S_{tot}=\frac{2\pi}{G}\frac{\dot H}{H^5}\left[\dot H+H^2-\frac{\gamma G\left(\Delta+2\right)}{2\pi}H^{2-\Delta}\right]
\,.
\ee
Notice that, for $\Delta=0$, the above expression
reproduces the usual BH entropy variation
$\dot S_{tot}=\frac{2\pi}{G}\frac{\dot H^2}{H^5}$, which takes
non-negative values for any $H$. This entails
the validity of the GSL of thermodynamics 
at any time within the standard framework. 
On the other hand, for $\Delta\neq0$, we infer
that $\dot S_{tot}\ge0$, provided that 
\be
\dot H>0\,\,\,\,\, \mathrm{and} \,\,\,\,\,
\left[\dot H+H^2-\frac{\gamma G\left(\Delta+2\right)}{2\pi}H^{2-\Delta}\right]\ge0\,,
\ee
or 
\be
\dot H<0\,\,\,\,\, \mathrm{and} \,\,\,\,\,
\left[\dot H+H^2-\frac{\gamma G\left(\Delta+2\right)}{2\pi}H^{2-\Delta}\right]\le0\,.
\ee
Therefore, we find that in our model the total entropy
variation is not necessarily a non-negative function, 
potentially yielding a violation of the GSL of thermodynamics.
A similar result is exhibited in the case of interacting Tsallis Holographic Dark Energy~\cite{Mamon:2020wnh} and BHDE in the absence of fluid radiation~\cite{Mamon:2020spa}.  A deeper understanding of this result requires further
study that will be presented elsewhere.

\section{Conclusions and outlook}
\label{Concl}
We have analyzed a generalized interacting
HDE model with Barrow entropy, which 
is based on a modified horizon endowed with a fractal
structure due to quantum gravitational corrections~\cite{BarrowBH}.
Within this framework, we have studied the evolution
of a spatially flat FLRW Universe filled by pressureless dark matter, radiation fluid and BHDE, whose 
IR cutoff is set equal to the Hubble length. 
We have assumed
a suitable interaction between the dark sectors
of the cosmos, which reproduces well-known 
interactions in the literature for some specific values of the model
parameters~\cite{Mamon:2020wnh}. The behavior of
various quantities, such as the BHDE density parameter, the equation of state parameter, the deceleration parameter, the jerk parameter and
the square of sound speed has been investigated. As a result, 
we have found that the present model
well describes the thermal history of the Universe, 
with the sequence of radiation, dark matter and dark energy
epochs, before resulting in a complete dark energy domination in the far future. Moreover, the cosmic evolution of the Universe is nontrivially
affected by both the interaction terms $b_1^2,b_2^2$ and 
Barrow parameter $\Delta$. 
Concerning the jerk parameter, we have shown that
$j>0$ and approaches to the $\Lambda$CDM model value $j=1$
as $z\rightarrow-1$. Finally, we have calculated the
square of sound speed to examine the classical stability
of our model against small perturbations. While in the absence
of radiation the interacting BHDE model is always
unstable ($v_s^2<0$), it becomes classically
stable ($v_s^2>0$) at early times when the effects or radiation
are taken into account. Results are summarized in Table~\ref{TabI}, 
from which we infer the constraints 
$0.03\le b_1^2\le0.15$, $0.01\le b_2^2\le0.08$ and $\Delta<0.55$ to allow consistency with observations. We notice that while
the bound on Barrow parameter is slightly wider than those
obtained in so-far literature~\cite{Anagnostopoulos:2020ctz,Leon:2021wyx,Barrow:2020kug,Jusufi:2021fek,Dabrowski:2020atl,Saridakis:2020cqq,Luciano:2022pzg,Vagnozzi:2022moj,DiGennaro:2022ykp}, constraints on the interaction terms are well-consistent or improve
past results (see for instance~\cite{Newp,Telep}).

 \begin{center}
\begin{table}[h]

   \hspace{-10mm} \begin{tabular}{c|c}
    \hline
 Without radiation  & With radiation   \\ 
 \hline
 \hline
  Quintessence/Cosmological constant ($z>-1$) & Quintessence/Cosmological constant ($z>-1$)   \\
  \vspace{-2mm}
  \\
       Phantom ($z\rightarrow-1$) &  Phantom ($z\rightarrow-1$) \\
       \vspace{-2mm}
       \\
            $-0.9\le\omega_{D_0}\le-0.64$ ($0.03\le b_1^2\le0.15, 0.01\le b_2^2\le0.08, \Delta=0.5$)  & $-0.9\le\omega_{D_0}\le-0.64$ ($0.03\le b_1^2\le0.15, 0.01\le b_2^2\le0.08, \Delta=0.5$)  \vspace{-1mm} \\ \\
           \hline
            \vspace{-1mm}
           \\
            $0.56\le z_t\le0.83$ ($0.04\le b_1^2\le0.07, 0.03\le b_2^2\le0.04, \Delta=0.5$)  & $0.52\le z_t\le0.82$ ($0.04\le b_1^2\le0.07, 0.03\le b_2^2\le0.04, \Delta=0.5$) \vspace{-1mm} \\ \\
           \hline
            \vspace{-1mm}
           \\
            $-0.46\le q_0\le-0.17$ ($b_1^2=0.05, b_2^2 =0.04, 0.4\le\Delta\le1$)  &  $-0.46\le q_0\le-0.17$ ($b_1^2=0.05, b_2^2 =0.04, 0.4\le\Delta\le1$)  \vspace{-1mm} \\ \\
           \hline
           \vspace{-1mm}
           \\
             $j>0$ (for all $z$), $j(z\rightarrow-1)\rightarrow1$  &  $j>0$ (for all $z$), $j(z\rightarrow-1)\rightarrow1$  \vspace{-1mm} \\ \\
            \hline
            \vspace{-1mm}
            \\
             $v_s^2<0$ (for all $z$) &  $v_s^2>0$ ($z\gg1$), $v_s^2<0$ ($z\rightarrow-1$) \\ \vspace{-1mm} \\
            \hline
    \end{tabular}
  \caption{Theoretical predictions of EoS parameter, transition redshift, deceleration parameter, jerk parameter and square of sound speed (for the current values of these quantities, we only report the best prediction of our models according to recent observations. Notice that such values are nearly equal for the two models, since effects of radiation fluid are only relevant at early times $z\gg1$).}
  \label{TabI}
\end{table}
\end{center}

As a further study, we have investigated the 
implications of gravity-thermodynamics in the BHDE model 
by assuming the apparent horizon as cosmological boundary. 
We have found
that, while in the standard HDE based on BH entropy ($\Delta=0$) 
the GSL of thermodynamics is always satisfied, on the other hand 
it might be
violated in BHDE ($\Delta\neq0$), depending on the evolution
of the Universe. 

Several aspects remain to be addressed. For instance, 
it would be interesting to investigate 
the properties of BHDE by considering
different IR cutoffs and/or interactions
between DM and DE. 
On the other hand, one might
think of extending the present analysis to the
case of HDE based on Kaniadakis entropy, 
which is a self-consistent extension of Boltzmann-Gibbs entropy that arises from relativistic statistical theory~\cite{Kana}. 
In this case, deviations from the standard framework are 
characterized
by the parameter $-1<K<1$.
Along this line, a challenging perspective is to search for any 
correspondence between the two generalizations
of HDE and possible relations between Barrow and Kaniadakis parameters. Finally, since our 
model provides a phenomenological attempt to 
account for quantum gravity effects in the evolution of the
Universe via the introduction of Barrow entropy, 
it is important to analyze whether our predictions reconcile with
more fundamental candidate theories  
or phenomenological model of quantum gravity. 
In this sense, preliminary attempts
to establish a connection between deformed entropies
and generalizations of the Heisenberg principle
induced by quantum gravitational effects have
been proposed in~\cite{Shababi,LucGUP,JizbaGUP}.
Work along these directions is in progress and is
reserved for future publications. 


\acknowledgments 

G.~G.~L. acknowledges the Spanish ``Ministerio de Universidades'' 
for the awarded Maria Zambrano fellowship and funding received
from the European Union - NextGenerationEU. He is also grateful for 
participation in the COST Association Action CA18108  ``Quantum Gravity Phenomenology in the Multimessenger Approach'' and LISA Cosmology Working group.  J. G. is partially supported by the Agencia
Estatal de Investigaci\`on grant PID2020-113758GB-I00 and an AGAUR
(Generalitat de Catalunya) grant number 2017SGR 1276.

\appendix
\section{Dark energy evolution}
\label{DEev}

In order to infer some more detail on the solution of Eq.~\eqref{Omp}, we express such equation as a differential system in the plane of the form
\be
\frac{d \Omega_D}{d z} \,=\, 3 \Delta \Omega_D \left[1 - \Omega_D - b_1^2 + (b_1^2 - b_2^2) \Omega_D \right]\,, \qquad
\frac{d a}{d z} \,=\, 2 - (2 - \Delta) \Omega_D\,.
\ee

A first integral associated to this system is now
\be
I\,=\, a^{-3 \Delta} \Omega_D^{-\frac{2}{b_1^2-1}} \left[1 - b_1^2 - \Omega_D + (b_1^2 - b_2^2) \Omega_D\right]^{\frac{(b_1^2-1) \Delta -2 b_2^2}{(b_1^2-1) (b_1^2 - b_2^2-1)}}\,,
\ee
and any solution of Eq.~\eqref{Omp} is given by setting $I=C$, where $C$ is an arbitrary constant. Specifically, in our case this constant can be fixed by imposing the condition $\Omega_D^0=0.73$. 

\bigskip

For Eq.~\eqref{Ompbis} we cannot proceed as above, since we are now in presence of an ordinary differential equation depending on two variables $\Omega_D$ and $\Omega_m$. However, we can consider the two following different regimes:
\begin{itemize}
\item $z\gg1$ and $z\rightarrow-1$, where we can roughly consider $\Omega_m\simeq0$ (see Fig.~\ref{Fig11}). In this case the differential system associated to Eq.~\eqref{Ompbis} is
\be
\frac{d \Omega_D}{d z} = 3 \Delta \Omega_D (4 - 4 \Omega_D - b_2^2 \Omega_D), \quad
\frac{d a}{d z} = 2 - (2 - \Delta) \Omega_D\,.
\ee
A first integral associated to this differential system is then
\[
I= a^{3 \Delta} \Omega_D^{-\frac{1}{2}} \left[(4 + b_2^2) \Omega_D-4\right]^{\frac{b_2^2 + 2 \Delta}{2b_2^2+8}}.
\]
\item On the other hand, we can specialize Eq.~\eqref{Ompbis}
to the present time $z=0$, where we know that $\Omega^0_m/\Omega^0_D\simeq1/3$. Substitution into Eq.~\eqref{Ompbis} then gives the differential system
\be
\frac{d \Omega_D}{d \tau} \,=\, 3 \Delta \Omega_D (4 - 5 \Omega_D  -b_1^2 \Omega_D/3-b_2^2 \Omega_D), \quad
\frac{d a}{d \tau} \,=\, 2 - (2 - \Delta) \Omega_D,
\ee
and the first integral associated to this system is
\be
I= a^{\Delta} \Omega_D^{-\frac{1}{6}} \{\left[b_1^2 + 3 (5 + b_2^2)\right]\Omega_D\}^{\frac{1}{6}+\frac{\Delta-2}{b_1^2 + 3 (5 + b_2^2)}}.
\ee
\end{itemize}

\end{document}